\documentclass[journal]{IEEEtran}
\IEEEoverridecommandlockouts
\usepackage{cite}
\usepackage{amsmath,amssymb,amsfonts}
\usepackage{algorithmic}
\usepackage{graphicx}
\usepackage{textcomp}
\usepackage{xcolor}
\usepackage{tikz}
\usepackage{booktabs}
\usepackage{array}
\usepackage{lscape}
\usetikzlibrary{shapes.geometric, arrows.meta, positioning}
\usepackage{placeins}
\usepackage{multirow}
\usepackage{tabularx}
\usepackage{booktabs}
\usepackage{caption}
\usepackage{array}
\usepackage{tablefootnote}
\usepackage{listings}
\usepackage{xparse}
\usepackage{float}
\usepackage{hyperref}
\usepackage{subfig}
\usepackage{enumitem}
\usepackage{forest}
\usepackage{makecell}
\usepackage{longtable}
\usepackage{booktabs}
\usepackage{hyperref}
\usepackage{enumitem}
\usepackage{soul}

\hypersetup{
  colorlinks=true,
  linkcolor=black,
  urlcolor=black,
  citecolor=black,
}
 
\usepackage{threeparttable}

 \newcommand{\filledcircle}{\tikz{\fill (0,0) circle (0.087cm);}}
\newcommand{\semicircle}{\tikz{\draw (0,0) circle (0.08cm); \fill[black] (0,0) -- (0,0.08cm) arc (90:270:0.08cm) -- cycle;}}
\newcommand{\emptycircle}{\tikz{\draw (0,0) circle (0.08cm);}}

\usetikzlibrary{arrows.meta, angles, quotes, shapes.geometric, positioning, fit, backgrounds}
\colorlet{linecol}{black!75}

\def\na{\textcolor{gray}{\textit{N/A}}}
\def\none{\textcolor{gray}{\textit{None}}}

\def\BibTeX{{\rm B\kern-.05em{\sc i\kern-.025em b}\kern-.08em
    T\kern-.1667em\lower.7ex\hbox{E}\kern-.125emX}}
\begin{document}


\title{Monitoring Vulnerabilities in Next-Generation Automotive Operating Systems}

\author{
    \IEEEauthorblockN{ Dimitri Simon\IEEEauthorrefmark{1}, Badis Hammi\IEEEauthorrefmark{1}, Joaquin Garcia-Alfaro\IEEEauthorrefmark{1}, Hervé Debar\IEEEauthorrefmark{1}\\ }
     \IEEEauthorblockA{\IEEEauthorrefmark{1}SAMOVAR, Télécom SudParis, Institut Polytechnique de Paris, 91120 Palaiseau, France\\
	dimitri.simon@telecom-sudparis.eu; badis.hammi@telecom-sudparis.eu \\
	joaquin.garcia\_alfaro@telecom-sudparis.eu; herve.debar@telecom-sudparis.eu\\}
    }

\maketitle

\begin{abstract}
Software-defined vehicles (SDVs) are revolutionizing transportation by integrating complex, interconnected hardware, and software systems. This evolution introduces significant security challenges. We present a comprehensive security analysis for SDVs, focusing on software vulnerabilities. We note that existing vulnerability assessment tools fall short in addressing operating systems vulnerabilities, particularly when it comes to efficiently analyzing diverse software stacks in realistic environments. We present and release a vulnerability assessment solution that efficiently addresses these limitations. Our approach combines systematic vulnerability discovery, leveraging public Common Vulnerabilities and Exposures (CVE) databases, within a dockerized development environment that evaluates exploitability risks. The results reveal both breadth of potential threats and the practical constraints we faced during exploitation. We discuss the implications for industry and research, and propose directions for building more resilient SDVs.\footnote{The VERA source code, along with the associated proof-of-concepts (PoCs) and exploit implementations, is publicly available in a dedicated repository at: \url{https://github.com/EternalDreamer01/vera}}

\end{abstract}


\section{Introduction}

The automotive industry is undergoing a radical transformation across both hardware and software domains. 
Architecturally, manufacturers are migrating from distributed legacy Electronic Control Units (ECU) toward zonal Electrical/Electronic (E/E) topologies and consolidated high-performance computers (HPCs) to accommodate the vastly increased data volumes generated by advanced sensors (e.g., cameras, radar, LiDAR, and so on) and domain controllers. Concurrently, the software landscape is shifting from proprietary, monolithic firmware toward the Software-Defined Vehicle (SDV)\footnote{A Software-Defined Vehicle (SDV) is a vehicle in which the majority of functionality is implemented, managed, and continuously enhanced through software, allowing features to evolve independently of fixed hardware. In SDVs, software spans the entire vehicle ecosystem, from infotainment and connectivity to safety-critical and autonomous driving functions, enabling scalability, rapid feature deployment, and lifecycle updates.} paradigm, where vehicles are increasingly built on open, Portable Operating System Interface (POSIX)-compatible platforms that support virtualization, containerization, and Over-The-Air (OTA) updates and software delivery, supplanting isolated vendor-specific stacks. 
This transition toward POSIX-compatible platforms has been driven by initiatives such as Automotive Grade Linux (AGL),\footnote{\url{https://www.automotivelinux.org/}} Android Automotive OS (AAOS),\footnote{\url{http://source.android.com/docs/automotive}} and Red Hat In-Vehicle OS (RHIVOS).\footnote{\href{https://www.redhat.com/en/blog/red-hat-vehicle-os-hardware-enablement-program}{https://www.redhat.com/en/blog}}

This convergence accelerates feature deployment and interoperability. more precisely, the adoption of POSIX-compliant architectures further enables portability across hardware platforms, simplifies integration of third-party software, facilitates reuse of open-source components, and streamlines development through standardised Application Programming Interfaces (APIs) and tooling. However, these advances also introduce new cybersecurity challenges, as increased connectivity, reliance on shared software layers, and broader ecosystem integration significantly expand the vehicle's attack surface \cite{Contextualizing_security_and_privacy_of_software_defined_vehicles_State_of_the_art_and_industry_perspectives}.

\begin{figure}[t!]
\begin{center}
\includegraphics[width=0.5\textwidth]{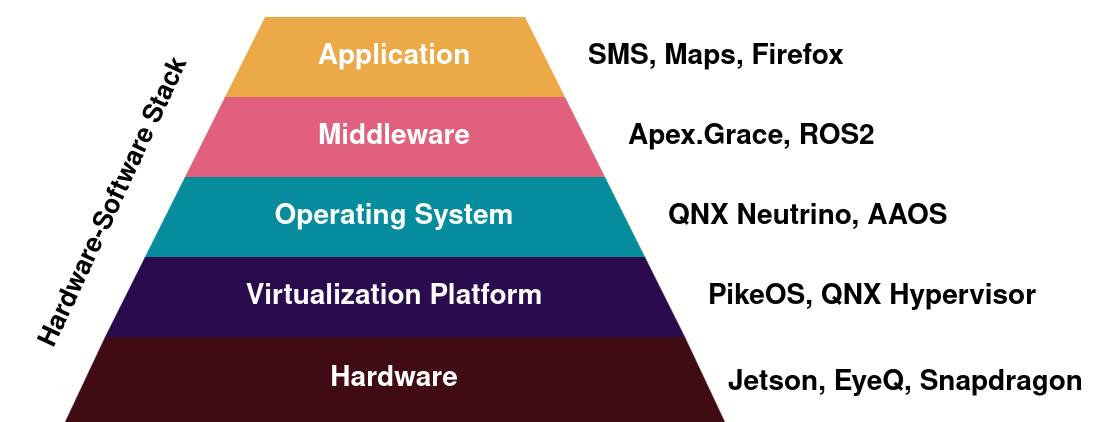}
 \caption{Hardware-software stack within next-generation SDVs}
 \label{fig:layers}
 \end{center}
 \end{figure}

Notably, security vulnerabilities in next-generation vehicles,\footnote{In the remainder of this paper, we use the term next-generation vehicle to refer to any vehicle that incorporates POSIX-compliant products or components.} including SDVs and non-SDV platforms that adopt POSIX-compliant components (e.g., infotainment systems), can emerge at any layer of the vehicle's stack, from application logic and middleware, through the operating system, hypervisor and drivers, down to firmware. Figure \ref{fig:layers} depicts this hardware-software stack within next-generation vehicles. 
The combination of increased software complexity, heterogeneous integration of legacy and modern components, ubiquitous connectivity (cellular, V2X, Wi-Fi, Bluetooth), and reliance on OTA updates substantially enlarges the adversary surface and enables novel attack vectors.

\subsection{Context and research questions}

POSIX-compatible operating systems and products have long accumulated large numbers of Common Vulnerabilities and Exposures (CVEs) in traditional computing and industrial contexts. The aforementioned adoption of POSIX architectures into SDVs will significantly expand traditional vehicle's attack surface~\cite{Contextualizing_security_and_privacy_of_software_defined_vehicles_State_of_the_art_and_industry_perspectives},   mainly due to increased connectivity, reliance on shared software layers, and broader ecosystem integration. Motivated by such potential threat, we investigate the following research questions: (1) to what extent do disclosed CVEs for POSIX-compatible ecosystems, including the virtualization, OSes, middleware and applications, actually affect vehicle software stacks; (2) whether those CVEs remain exploitable in the constrained and Original Equipment Manufacturer (OEM)-hardened environments of modern vehicles; and (3) how successful exploitation would affect vehicle safety and security given the highly interconnected architecture of automotive platforms.
Answering  these research questions requires moving beyond catalogue-style vulnerability listings to measured, penetration-testing-driven validation across software layers. 
This paper is driven by that objective and aims to empirically assess the presence, exploitability, prerequisites, and impact of POSIX CVEs in next-generation vehicles contexts.

\subsection{Research gaps and motivation}

\textbf{Uncertainty about the relevance and exploitability of POSIX/Linux CVEs in vehicle contexts:}
the automotive industry is actively migrating toward POSIX-based platforms (e.g., AGL, Android Automotive variants, Red Hat In-Vehicle OS), which exposes vehicle software to a broader ecosystem of libraries and kernel surfaces. At the same time, POSIX/Linux ecosystems continue to generate large numbers of CVEs \cite{Android_Automotive_OS_Update_Bulletin_September_2024}, with a recent surge in reported kernel CVEs,\footnote{\href{https://tuxcare.com/blog/the-linux-kernel-cve-flood-continues-unabated-in-2025/}{https://tuxcare.com/blog/}} raising the question whether vulnerabilities discovered in traditional computing environments are present, exploitable, or impactful in next-generation vehicles' configurations. 
The mapping from a disclosed CVE to a practical, vehicle-level exploit is nontrivial and under-explored \cite{Harness_Transparent_and_Lightweight_Protection_of_Vehicle_Control_on_Untrusted_Android_Automotive_Operating_System}. 

\textbf{Tooling limitations \textcolor{black}{to inspect/parse reports}:} 
existing vulnerability scanners and frameworks were largely designed for 
\textcolor{black}{scanning, without considering parsing (filter, sort, merge reports) and the relevance of a CVE (available exploits, execution conditions for each exploit, type e.g, RCE, LPE, etc.).
Recent evaluations of container and image scanners \cite{Vexed_by_VEX_tools_Consistency_evaluation_of_container_vulnerability_scanners}\cite{DAVS_Dockerfile_Analysis_for_Container_Image_Vulnerability_Scanning} demonstrate this inconsistent coverage.
}

\textbf{Shortage of penetration-testing–driven, cross-layer exploitability studies for next-generation vehicles:}
while the automotive security literature contains numerous component-focused analyses (ECU, firmware auditing, telematics investigations) \cite{Real_time_security_warning_and_ECU_identification_for_in_vehicle_networks, Review_of_the_security_of_backward_compatible_automotive_inter_ECU_communication, An_endogenous_security_study_of_telematics_box_in_intelligen_connected_vehiclesc}, comprehensive studies that combine systematic vulnerability discovery with end-to-end penetration testing across the full software stack of next-generation vehicles (applications → middleware → OS/hypervisor → firmware → boot) remain scarce \cite{Applying_security_testing_techniques_to_automotive_engineering, ICVTest_A_Practical_Black_Box_Penetration_Testing_Framework_for_Evaluating_Cybersecurity_of_Intelligent_Connected_Vehicles, Systematic_Risk_Analysis_of_Multi_Stage_Attacks_in_Zonal_Automotive_E_E_Architecture}. Hence, there is a need for realistic testing in next-generation vehicles' development lifecycles and practical frameworks to evaluate real exploitability rather than purely static enumeration \cite{Cybersecurity_testing_for_automotive_domain_A_survey}.


\subsection{Contributions of this work}

This study addresses the identified research gaps by:
\begin{enumerate}
\item Providing a comprehensive survey of automotive operating systems, middleware, and deployment architectures, with a systematic mapping of platforms to roles, common use cases, and assurance constructs.
\item Providing an empirical assessment of how POSIX/Linux CVEs propagate into vehicle stacks and their practical exploitability via controlled penetration tests.
\item Releasing a vulnerability scanning framework \textcolor{black}{more specifically designed for automotive POSIX-based operating systems}. 
\item Integrating a static vulnerability discovery with controlled penetration-testing validation to evaluate the practical exploitability of CVEs across software layers.
\item Releasing experimental artifacts and results to foster reproducibility, transparency, and comparative analysis in future research.
\end{enumerate}


\section{Background and systematization of SDV components and security}

\subsection{Standardized frameworks for vulnerability identification and risk assessment}
\label{sec:standards}

Disclosed vulnerabilities are catalogued in public repositories such as Common Vulnerabilities and Exposures (CVE), US National Vulnerability Database (NVD), and EU Vulnerability Database (EUVD).\footnote{\scriptsize{\url{http://cve.org}, \url{http://nvd.nist.gov}, \url{http://euvd.enisa.europa.eu/}}} Vulnerability scanners report vulnerabilities with their \emph{status} on the target system (e.g., fixed, not-fixed, and unknown).
This feature is called tracking mechanism: the scanner links software artifacts (packages, libraries, commits, or build recipes) to known vulnerabilities over time, typically by maintaining mappings between component identifiers/versions and CVE entries, and updating these mappings as new vulnerabilities and patches are published.
This information helps assess the effective exposure of the system and reduces false positives.
However, there is no reliable mechanism to track installed patches or software updates, making it impossible to accurately determine the true remediation state of vulnerabilities on the target system.

The severity of an identified vulnerability is typically expressed using the Common Vulnerability Scoring System (CVSS), which produces a standardized numeric score on a 0-10 scale: Low: 0 -- 3.9; Medium: 4 -- 6.9; High: 7 -- 8.9; and Critical: 9 -- 10.  Different vendors or authorities may assign distinct CVSS vectors for the same vulnerability, which motivates the common practice of reporting the maximum observed score across trusted sources.

\begin{table*}[!ht]
	\centering
	\begin{threeparttable}
	\scriptsize
	
	\setlength{\extrarowheight}{2pt}
	
	\caption{Standards commonly applied to operating systems and platforms in next-generation vehicles}
	\label{tab:standards_simple}
	\begin{tabular}{|p{1.75cm}|p{1.7cm}|p{2.7cm}|l|p{2.8cm}|p{1.65cm}|p{3.2cm}|}
		\hline
		\textbf{Standard} & \textbf{Scope / Target}		& \textbf{Typical output / assessment}                              & \textbf{Levels} & \textbf{Relevance to SDV}                               & \textbf{Primary focus}         & \textbf{Regulatory status}                                                                    \\
		\hline
		ISO 26262		& Road vehicles					& ASIL assignment, safety concept, FMEDA,\protect\footnotemark HARA,\protect\footnotemark \newline V\&V\protect\footnotemark evidence & ASIL A--D       & High -- safety-critical ECUs and system-level safety                            & Functional safety              & Widely adopted industry standard; commonly referenced by OEMs and regulators (not a law)      \\
		\hline
		ISO/PAS 8926	& Road vehicles & Guidance for integrating legacy/third-party software into ISO 26262 workflows & \na & High - reuse of libraries and legacy modules in SDV stacks & Software-integr. cybersecurity & Informative guidance; complements ISO 26262 \\
		\hline
		ISO/SAE 21434 	& Road vehicles					& CAL, cybersecurity threats in vehicles across their lifecycle. 	& CAL 1--4	& High -- Cybersecurity threats in vehicles 					& Process-oriented cybersecurity 	& Standard widely adopted ; CAL certification less exploited \\
		\hline
		IEC 61508		& Generic E/E /PE\protect\footnotemark systems across industries & SIL requirements, validation and verification evidence	& SIL 1--4	& Medium -- useful when porting industrial modules or using industrial components & Functional Safety	& International standard used across industries; often referenced in industrial contexts        \\
		\hline
		ISO/IEC 15408-5	& IT products and components	& EAL, protection profiles, certification reports					& EAL 1--7	& Medium -- security assurance for OS components and modules	& Security							& Recognized certification framework; adopted by national schemes for procurement and assurance \\
		\hline
	\end{tabular}
	\end{threeparttable}
\end{table*}
\footnotetext[15]{Hazard Analysis and Risk Assessment}
\footnotetext[16]{Verification and Validation}
\footnotetext[17]{Failure Modes, Effects, and Diagnostic Analysis}
\footnotetext[18]{Programmable Electronic}

To assist prioritization, practitioners increasingly use the Exploit Prediction Scoring System (EPSS), which provides a probabilistic estimate (0--1) of the likelihood that a given vulnerability will be exploited in the wild within the next 30 days. EPSS complements CVSS by adding a data-driven notion of exploit likelihood to severity metrics. Vulnerability and product records are correlated using standardized naming such as the Common Platform Enumeration (CPE),\footnote{\url{https://nvd.nist.gov/products/cpe}} which reduces ambiguity when mapping CVEs to specific products, versions, and packages. 

Prior empirical work has shown that severity scores alone are an imperfect proxy for real-world exploitability and risk. For example, studies comparing CVSS and observed exploitation find only limited predictive power, underscoring the need to combine severity, exploit-likelihood (EPSS), provenance, and contextual factors when prioritizing vulnerabilities~\cite{Allodi2014}.

In addition to CVSS and EPSS,\footnote{
\url{https://www.first.org/epss}} practitioners often consider qualitative factors such as the presence of active real world exploitations. This can be known from (1) the Known Exploited Vulnerabilities Catalog (KEV Indicator/Catalog)\footnote{\url{https://www.cisa.gov/known-exploited-vulnerabilities-catalog}} and (2) the availability of exploit code, commonly tracked using threat intelligence feeds and databases (e.g., Exploit Database (ExploitDB)\footnote{\url{https://www.exploit-db.com/}} and GitHub repositories hosting proof-of-concept exploits).

During our study, we particularly considered broad critical classes of vulnerabilities and attack vectors that represent major threats to next-generation vehicle systems regarding passengers' safety and privacy, as listed below:
\begin{description}
	\item[Remote Code Execution (RCE):] a vulnerability that allows an attacker to execute arbitrary code on a target system remotely, often without prior authentication, potentially leading to full system compromise.
	\item[Local Privilege Escalation (LPE):] sometimes referred as Elevation of Privilege (EoP), is a flaw that enables a local user or process to elevate privileges, bypassing restrictions on access to sensitive data or administrative actions. Such vulnerabilities can turn limited compromises into complete system control.
	\item[Sandbox or Virtual Machine Escape (SBX/VME):] a vulnerability that permits code execution or data access on the host machine from within a sandboxed environment (e.g., browser sandboxes, Docker containers, or virtual machines), undermining isolation guarantees.
\end{description}

\subsection{Safety and security standards}

Evaluating operating system security in automotive and Advanced Driver-Assistance Systems (ADAS) platforms often involves compliance with international safety and security standards from ISO and IEC. These standards provide structured methodologies for assessing risk and ensuring that systems meet rigorous safety and assurance requirements. Table \ref{tab:standards_simple} provides a summary of the main standards.

\textbf{ISO 26262} \cite{ISO_26262} is a functional safety standard designed specifically for road vehicles. It introduces \textit{Automotive Safety Integrity Levels (ASIL)} from level A to level D, where ASIL A represents minimal risk mitigation requirements and ASIL D denotes the highest safety rigor and most stringent engineering measures.

\textbf{ISO/PAS 8926} \cite{ISO_PAS_8926} extends ISO 26262 by offering guidance on integrating pre-existing software components or architectural elements not originally developed under ISO 26262. It addresses challenges of reusing legacy or third-party software within safety-critical automotive systems.

\textbf{ISO/SAE 21434} \cite{ISO_SAE_21434} is a process-oriented standard for automotive cybersecurity.
It specifies the level of rigor required in cybersecurity activities throughout the vehicle's lifecycle.
It introduces \textit{Cybersecurity Assurance Level (CAL)} from level 1 to level 4, where CAL 1 represents basic cybersecurity design and CAL 4 the highest level of cybersecurity measures.
The CAL guides the cybersecurity activities from concept through development, production, operation, maintenance, and decommissioning.

\textbf{IEC 61508} \cite{IEC_61508} is a generic cross-industry functional standard for electrical, electronic, and programmable electronic safety systems. It defines \textit{Safety Integrity Levels} (SIL 1-4), with SIL 1 indicating basic risk reduction measures and SIL 4 representing the highest level of risk mitigation.

\textbf{ISO/IEC 15408-5 (Common Criteria)} \cite{ISO_IEC_15408_5} specifies \textit{Evaluation Assurance Levels} (EAL 1--7), which indicate the depth and rigor of security evaluation performed on IT products. Higher EALs require more extensive documentation and testing but do not inherently guarantee higher security, only greater assurance through structured evaluation.

Table~\ref{tab:os_certifications} summarizes prominent automotive virtualization platforms, operating systems and middleware together with their security/safety posture based on their formal certification levels. 
Notably, several widely adopted platforms like AGL, Eclipse S-CORE, and Zephyr, do not currently possess any recognized formal automotive certification.
Hardware platforms can also be certified. For example, the Snapdragon 855 (SM8150) platform has been certified EAL 4+.

\begin{table}[h]
	\centering
	\scriptsize
	\setlength{\tabcolsep}{4pt}
	\setlength{\extrarowheight}{2pt}
	
	\caption{Automotive OS, middleware and hypervisors: EAL (ISO/IEC 15408-5), ASIL (ISO 26262), SIL (IEC 61508) and CAL (ISO/SAE 21434) status.}\label{tab:os_certifications}
	\begin{tabular}{b{3.0cm} m{1.9cm} b{0.5cm} p{0.5cm} p{0.5cm} p{0.5cm}}
		Hypervisor / OS / Middleware & Vendor             & EAL   & ASIL  & SIL   & CAL   \\ \hline
		QNX Hypervisor  & \multirow{2}{*}{BlackBerry}         & \none & D     & 3     & 2     \\
		QNX Neutrino    &                    & 4+    & D     & 3     & 2     \\ \hline
		VxWorks 7       & \multirow{3}{*}{Wind River}         & 6+    & D     & 3     & \none \\
		Wind River Helix &                   & \none & D     & 3     & \none \\
		Wind River Linux &                   & \none & \none & \none & \none \\ \hline
		INTEGRITY Multivisor & {Green Hills}   & \none & D     & 3     & \none \\
		INTEGRITY       &                    & 6+    & D     & 3     & 4     \\ \hline
		PikeOS Hypervisor & \multirow{2}{*}{SYSGO}            & 5+    & D     & 3     & \none \\
		PikeOS Native   &                    & 5+    & D     & 3     & \none \\ \hline
		LynxSecure      & Lynx Software Technologies & \none & D     & 3 & \none \\
		LynxOS          &                    & \none & D     & 3 & \none \\ \hline
		Mentor Embedded Hypervisor & \multirow{2}{*}{Siemens} & \none & D     & 3     & \none \\
		Mentor Nucleus  &                    & \none & D     & 3     & \none \\ \hline
		COQOS Hypervisor & OpenSynergy       & \none & D     & 3     & \none \\ \hline
		DRIVE OS 6      & \multirow{2}{*}{NVIDIA}             & \none & D     & \none & 2*    \\
		DRIVE OS 5      &                    & \none & B     & \none & 2*    \\ \hline
		RHIVOS          & Red Hat            & \none & B     & \none & \none \\ \hline
		Eclipse S-CORE  & Eclipse Foundation & \none & \none & \none & \none \\ \hline
		Zephyr          & Zephyr Project     & \none & \none & \none & \none \\ \hline
		Ubuntu 18.04    & Canonical          & 2+    & \none & \none & \none \\ \hline
	\end{tabular}
	\parbox{\linewidth}{\scriptsize
		~~\\
		*NVIDIA DRIVE OS is certified to ISO 21434 via QNX.
	}
\end{table}

\subsection{Automotive OS and middleware platforms}

\begin{table}[]
	\setlength{\tabcolsep}{2.2pt}
    \color{black}
	\centering
	\scriptsize
	\caption{Vehicle models, their IVI/telematics OS and SoC, and safety/ADAS OS by year of any change.}
	\label{tab:mapping_cars}
	\begin{tabular}{lll p{1.7cm} p{2.8cm}}
		\hline
		Year  & Brand          & SoC     & IVI/Telem. OS        & S/A. OS                           \\
		\hline

2014  & Tesla          & -       & Linux                & -                                 \\
2015  & Honda          & -       & Android 4.0          & DRIVE OS                          \\
2016  & Cadillac       & -       & AGL; AAOS            & QNX                               \\
2017  & Tesla          & -       & Linux                & -                                 \\
2019  & Hyundai        & -       & AGL                  & DRIVE OS; QNX                     \\
2019  & Tesla          & -       & Linux                & -                                 \\
2020  & General Motors & A3960   & AGL; AAOS 29         & VxWorks; QNX                      \\
2020  & Mitsubishi     & SM8150  & AGL                  & DRIVE OS; VxWorks; QNX            \\
2020  & Polestar       & A3960   & AAOS                 & DRIVE OS; QNX                     \\
2020  & XPeng          & -       & Linux                & DRIVE OS; QNX                     \\
2020  & Volvo          & A3960   & AAOS                 & DRIVE OS; QNX                     \\
2020  & Volkswagen     & -       & AGL; AAOS            & DRIVE OS; QNX                     \\
2023  & Renault        & SM8150  & AGL; AAOS 32         & QNX                               \\
2021  & Ford           & -       & AGL; AAOS            & DRIVE OS; INTEGRITY; VxWorks; QNX \\
2021  & Lucid Motors   & -       & AAOS                 & DRIVE OS; QNX                     \\
2021  & Mercedes-Benz  & -       & AGL                  & DRIVE OS; QNX                     \\
2021  & NIO            & -       & AAOS                 & DRIVE OS; QNX                     \\
2022  & Mitsubishi     & SA8195P & AGL; AAOS 32         & DRIVE OS; INTEGRITY; VxWorks; QNX \\
2022  & Nissan         & SA8155P & AGL; AAOS 32         & DRIVE OS; QNX                     \\
2022  & Subaru         & SA8155P & AGL; AAOS            & QNX                               \\
2024  & BYD            & SA8155P & AAOS 34              & DRIVE OS; QNX                     \\
2022  & XPeng          & -       & Linux                & DRIVE OS; QNX                     \\
2022  & Toyota         & SA8155P & AGL                  & DRIVE OS; INTEGRITY; VxWorks; QNX \\
2023  & General Motors & SA8155P & RHIVOS; AGL; AAOS 32 & VxWorks; QNX                      \\
2023  & Honda          & SA8155P & AGL; AAOS 32         & DRIVE OS; QNX                     \\
2023  & Tesla          & -       & Linux                & -                                 \\
2023  & Polestar       & SA8155P & AAOS 32              & DRIVE OS; QNX                     \\
2023  & Porsche        & -       & AAOS                 & DRIVE OS; QNX                     \\
2023  & BMW            & SA8155P & AGL; AAOS            & DRIVE OS; VxWorks; QNX            \\
2022  & Hyundai        & -       & AAOS                 & QNX                               \\
2023  & Polestar       & SA8155P & AAOS 33              & DRIVE OS; QNX                     \\
2024  & General Motors & SA8195P & RHIVOS; AGL; AAOS 34 & VxWorks; QNX                      \\
2024  & Volvo          & SA8155P & AGL; AAOS            & DRIVE OS; QNX                     \\
2025  & NIO            & -       & AAOS                 & QNX                               \\
2025  & Renault        & SM8150  & RHIVOS; AGL; AAOS 32 & QNX                               \\
2026* & Cadillac       & -       & RHIVOS; AGL; AAOS    & QNX                               \\
2026* & General Motors & -       & RHIVOS; AGL; AAOS    & DRIVE OS; VxWorks; QNX            \\
2026* & Tesla          & -       & Linux                & -                                 \\
		\hline
	\end{tabular}
	\parbox{\linewidth}{\scriptsize
		~~\\
		\textbf{Legend ---}
		\textbf{SoC:} Infotainment SoC (System-on-Chip).
		\textbf{IVI/Telem. OS:} Infotainment and Telematics OS.
		\textbf{S/A. OS:} Safety/ADAS OS.
		\textbf{Linux:} Linux-based.
	}
\end{table}

\begin{table*}[ht]
	\centering
	\scriptsize
	\caption{Virtualization platforms, OS and middleware preferred use cases.}
	\label{tab:vp_os_mw_use_cases}
	\bgroup
    \def\arraystretch{1.25}

	\begin{tabular}{r|c|rlccccc}
		\multirow{2}*{\textbf{Vendor}} & \multirow{2}*{\textbf{Virtualization Platform}} & \multicolumn{7}{c}{\textbf{OS \& Middleware Purposes}} \\
		~ & ~ & \textbf{OS} & \textbf{Middleware} & Safety & AI inf. & Info. & Telem. & V2X, Connect. \\
		\hline

		Black Berry & QNX Hypervisor & QNX Neutrino & & \checkmark & & & \checkmark & \checkmark \\ \hline
		\multirow{2}*{Wind River} & \multirow{2}*{Wind River Helix} & VxWorks & & \checkmark & & & & \checkmark \\ \cline{3-9}
		& & WindRiver Linux & & \checkmark & & & \checkmark & \checkmark \\ \hline
		\multirow{2}*{Apex.AI} & & Apex.OS & & \checkmark & & & & \checkmark \\ \cline{3-9}
		& & & Apex.Grace & & \checkmark & & & \\ \hline
		NVIDIA & & NVIDIA DRIVE OS & & & \checkmark \\ \hline
		Green Hills & INTEGRITY Multivisor & INTEGRITY & & \checkmark & & & & \checkmark \\ \hline
		SYSGO & PikeOS Hypervisor & PikeOS Native & & \checkmark & & & \checkmark & \checkmark \\ \hline
		Siemens & Mentor Embedded Hypervisor & Nucleus & & \checkmark & & & & \checkmark \\ \hline
		Lynx Software Technologies & LynxSecure & LynxOS & & \checkmark & & & & \checkmark \\ \hline
		OpenSynergy & COQOS Hypervisor & & \begin{minipage}[l]{2cm}
			\makecell{Blue SDK\\RapidLaunch SDK\\Radio Tuner SDK}
		\end{minipage} & & & \checkmark & \checkmark & \checkmark \\ \hline
		Red Hat & & RHIVOS & & \checkmark & & \checkmark & \checkmark & \checkmark \\ \hline
		Google & & AAOS & & & & \checkmark & \checkmark & \checkmark \\ \hline
		Linux Foundation & & AGL & & & & \checkmark & \checkmark & \\ \hline
		Eclipse Foundation & & & Eclipse S-CORE & \checkmark & & & \checkmark & \checkmark \\ \hline
		Zephyr Project & & Zephyr & & \checkmark & & & & \\ \hline
		\multirow{2}*{AUTOSAR Foundation} & & & AUTOSAR Classic & \checkmark & & & & \checkmark \\ \cline{3-9}
		& & & AUTOSAR Adaptive & \checkmark & & & \checkmark & \checkmark \\ \hline
		Open Robotics & & & ROS2 & & \checkmark & & & \checkmark \\ \hline
	\end{tabular}
	\egroup
	
	\parbox{\linewidth}{\scriptsize
		~~\\
		\textbf{Legend ---}
		\textbf{AI inf.:} AI inference
		\textbf{Info.:} Infotainment
		\textbf{Telem:} Telematics
		\textbf{V2X, Connect.:} V2X Connectivity
	}
\end{table*}


It is important to clearly distinguish middleware solutions from operating systems, as these two software layers fulfill fundamentally different roles within the vehicle software stack. Middleware provides runtime services and communication abstractions between the OS and applications (e.g., message buses, Remote Procedure Calls (RPC), and lifecycle management), whereas the OS is responsible for hardware abstraction, resource management, process scheduling, and low-level services. Additionally, a virtualization layer may be deployed above the hardware to host and isolate multiple operating systems, thereby enhancing security and system robustness.

Generally, virtualization platforms, operating systems and middleware are typically combined to leverage complementary strengths (e.g., real-time determinism, isolation, rich application ecosystems, or small-footprint operation) and to satisfy the varied requirements of infotainment, telematics, ADAS, and safety-critical subsystems. Table \ref{tab:vp_os_mw_use_cases}  summarizes the preferred use cases of the main platforms.

In Table \ref{tab:mapping_cars}, we present a list of vehicle models alongside the OSes and middleware they implement. Additional details regarding their specific use cases are provided in Table~\ref{tab:vp_os_mw_use_cases}. Mapping these components to the architecture illustrated in Figure~\ref{fig:layers}, we classify them as follows:
\begin{itemize}
\item Systems-on-Chip (SoCs) is the hardware layer.
\item QNX Hypervisor, Wind River Helix, and PikeOS Hypervisor serve as virtualization platforms.
\item QNX Neutrino, INTEGRITY, and VxWorks are real-time operating systems (RTOS) that support virtualization and provide middleware integrations, effectively spanning the virtualization, OS, and middleware layers.
\item AAOS, AGL, DRIVE OS, and RHIVOS are operating systems that incorporate middleware capabilities, bridging the OS and middleware layers.
\item ROS2, AUTOSAR, and Eclipse S-CORE operate strictly at the middleware layer.
\end{itemize}


To our knowledge, our work is the first to systematically present software‑stack component certifications used within SDVs, and to determine and report vehicle models' SoCs and operating systems based on company disclosures.
Although it is not exhaustive, this table offers a snapshot of the diverse platforms currently used in the automotive sector. The challenge in compiling this information lies in the proprietary nature of automotive systems, as manufacturers often keep details about their OS and middleware configurations confidential. Furthermore, frequent updates to software stacks, customizations per model, and the lack of standardized reporting make it difficult to provide a comprehensive overview.
We quickly present the benchmarked systems.

\textbf{Eclipse S-CORE}\footnote{\url{https://eclipse-score.github.io}} is an open-source initiative that aims to provide a safe core stack (middleware) for software-defined vehicles and explicitly targets safety-critical in-vehicle domains such as ADAS. The project is actively developing safety artefacts and has been the subject of recent independent safety assessments, but there is no publicly documented ISO 26262 certification at this time.

\textbf{Zephyr}\footnote{\url{https://zephyrproject.org/safety-and-zephyr-rtos/}} is a real-time operating system originally designed for resource-constrained devices. Recently, it has attracted interest for use in next-generation vehicle subsystems (e.g., low-latency controllers and safety-adjacent components).
While not yet broadly certified for automotive use, there are ongoing efforts in the community and industry to align Zephyr with automotive safety and dependability goals (including workstreams towards ISO 26262 and IEC 61508 conformance)\cite{Zephyr_in_Automotive_and_SDV}.

\textbf{Red Hat In-Vehicle OS (RHIVOS)}\footnote{\url{https://www.redhat.com/en/resources/in-vehicle-operating-system-detail}} is positioned as a mixed criticality platform for both safety-critical applications like ADAS and non-safety critical like infotainment.
In practice, early deployments and discussions emphasize its use in infotainment and application hosting, although RHIVOS is explicitly intended to support a wider set of vehicle functions through the AutoSD initiative.\footnote{\url{https://sigs.centos.org/automotive/about/}}

\textbf{VxWorks}\footnote{\url{https://www.windriver.com/products/embedded/vxworks}} is a real-time operating system (RTOS) designed for embedded and safety-critical systems. VxWorks modularizes core services into isolated components, enabling configurable footprints and easier updates. It supports multi-core SMP, mixed-criticality applications, POSIX APIs, and advanced networking and security stacks. Deterministic scheduling, low-latency interrupt handling, and certified safety/security extensions make it common in aerospace, defense, industrial control, medical devices, and sometimes in vehicles.

\textbf{QNX Neutrino}\footnote{\url{https://www.qnx.com/products/intl/neutrino_rtos/}} is a microkernel-based RTOS built for reliability and fault isolation in embedded systems. Its minimal microkernel provides IPC, scheduling, and interrupt management; drivers and services run in user space, reducing system-wide crashes. It offers POSIX compatibility, deterministic real-time performance, and strong filesystem/network stacks. Widely used in automotive infotainment, industrial automation, and safety-critical domains, QNX emphasizes modularity, fast boot, and system resiliency.

\textbf{TeslaOS}\footnote{\url{https://github.com/teslamotors/linux}} -- the publicly shared ``TeslaOS" materials primarily refer to Tesla's infotainment/system Linux stack (NVIDIA Tegra–based) used for media, navigation, UI, and vehicle-connected services. It includes kernel sources, drivers, and user-space components for the center display and media domain, with OTA updates and tight hardware integration. The proprietary vehicle-control and autonomy software remains separate and is not included in the public infotainment repository.

\textbf{Android Automotive OS (AAOS)}\footnote{\url{https://developer.android.com/training/cars/platforms/automotive-os}} is Google’s in-vehicle operating system version built from Android to run directly on car hardware. It provides native vehicle integration for instrument clusters, media, navigation, and voice assistant, plus access to the Android app ecosystem tailored for automotive use.

\textbf{Automotive Grade Linux (AGL)}\footnote{\url{https://www.automotivelinux.org}} is an open-source Linux-based software stack for automotive applications, collaborative-developed by OEMs and suppliers. AGL provides reference implementations for infotainment, telematics, instrument clusters, and domain controllers using a common middleware layer, flexible app framework, and standardized APIs. Its open governance accelerates integration, portability across hardware, and ecosystem innovation, enabling manufacturers to build customizable, connected vehicle features while reducing duplicated engineering effort.

\textbf{ROS 2}\footnote{\url{https://www.ros.org}} is a modular, open-source robotics middleware framework designed for distributed, real-time-capable robotic systems. It provides publish/subscribe communication, lifecycle management, and QoS-configurable DDS-based transports for reliable inter-process messaging across heterogeneous hardware. ROS 2 emphasizes portability, security, and deterministic behavior for production use, with rich libraries for perception, planning, and control. It is widely adopted in research, industrial robots, and autonomous vehicles for integrating sensors, algorithms, and system orchestration.

We can pragmatically distinguish platforms used for safety/ADAS from those primarily intended for infotainment:
\begin{itemize}
	\item \textbf{Safety-critical platforms:}
		QNX Neutrino, VxWorks, INTEGRITY, and Red Hat In-Vehicle OS (RHIVOS) are Real-Time Operating Systems (RTOS) or seperation kernels that are widely used for safety-critical functions.
		PikeOS Hypervisor \& Native in particular is a hypervisor and RTOS designed to consolidate multiple subsystems with strong isolation guarantees.
		NVIDIA DRIVE OS is a vehicle compute platform used for AI inference workloads. It is also often positioned with safety considerations for those domains.
	\item \textbf{Non-safety-critical platforms:} Automotive Grade Linux (AGL), TeslaOS, Android Automotive OS (AAOS) and other Android variants are commonly adopted for infotainment, telematics, and rich application hosting (user experience, media, app ecosystems).
\end{itemize}

\subsection{Automotive OS deployment}

Building upon the operating systems and middleware platforms introduced in Section II.C, this subsection details how these specific components are practically deployed and isolated across the lower layers of the architectural stack presented in Figure \ref{fig:layers}.
\begin{figure*}[t!]
	\centering
	\subfloat[]{\includegraphics[width=5cm]{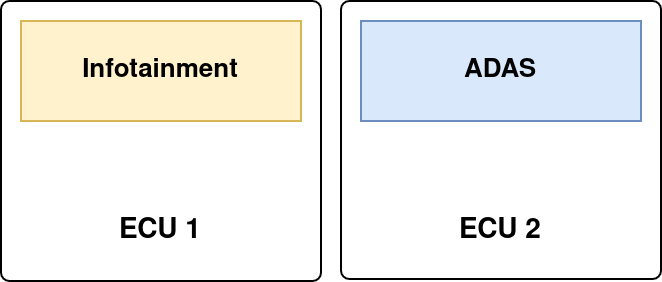}}
	\hspace{0.2cm}
	\subfloat[]{\includegraphics[width=5cm]{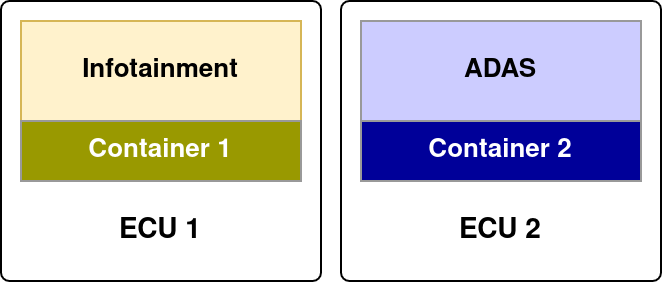}}
	\hspace{0.2cm}
	\subfloat[]{\includegraphics[width=5cm]{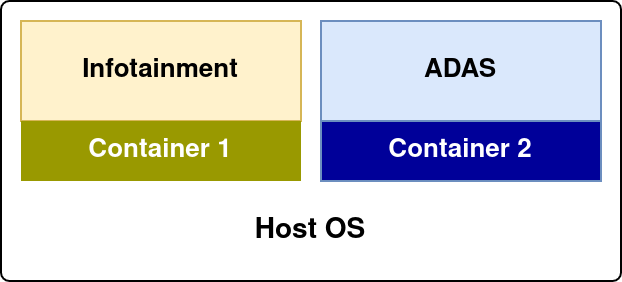}}
	\caption{Deployment architectures for infotainment and ADAS: (a) physical isolation (per-ECU), (b) physical isolation with containerization, (c) virtualized isolation using hypervisors/containers with software/hardware partitioning.}
	\label{fig:os_deployment}
\end{figure*}
Because modern automotive platforms rely on a complex mix of virtualization technologies, operating systems, and middleware, robust deployment architectures are essential. In Figure \ref{fig:os_deployment}, we illustrate three canonical deployment architectures for hosting infotainment and ADAS workloads.These architectures demonstrate how the foundational hardware and OS layers from Figure \ref{fig:layers} are configured in next-generation vehicles to embed and isolate multiple operating systems securely. 
To demonstrate how these distinct architectural layers are implemented in practice, Table \ref{tab:mapping_cars}  details the hardware and operating system combinations utilized in real-world vehicles, while Table\ref{tab:vp_os_mw_use_cases}  categorizes the preferred use cases across the virtualization, OS, and middleware platforms.
High Performance Computer (HPC) is typically the Host OS, and the ADAS platform might be directly on the same Electronic Control Unit (ECU) as the HPC whereas Infotainment runs in a container.
Next-generation vehicles typically combine hardened RTOSes or separation kernels for safety-critical subsystems with Linux/Android-based platforms for infotainment, connected via middleware standards to meet mixed-criticality, performance and ecosystem requirements.


\begin{itemize}
	\item \textbf{Physical isolation (cf. Figure \ref{fig:os_deployment}.a):} distinct ECUs or separate compute platforms host different domains (e.g., infotainment versus ADAS). Isolation is enforced by hardware separation and dedicated Input/Output, which simplifies safety and security arguments and reduces cross-domain attack surface at the cost of increased weight, cost and reduced resource sharing.
	\item \textbf{Physical isolation with containerization (cf. Figure \ref{fig:os_deployment}.b):} still distinct ECUs or separate compute platforms for different domains. However, the containerization adds a security layer.
	\item \textbf{Virtualized isolation (cf. Figure \ref{fig:os_deployment}.c):} multiple domains are consolidated onto a single hardware platform using hypervisors or container runtimes. Logical partitions (VMs or containers) present independent execution environments while sharing the same physical resources. This architecture improves resource use and maintainability but introduces evasion risks, misconfiguration, and inter-partition side channels.
\end{itemize}

Although consolidation reduces cost and improves agility, it also amplifies attack surface and cross-domain impact (e.g., sandbox or hypervisor escape, cross-partition escalation, and so on).
Empirical automotive security research has repeatedly illustrated how shared or exposed subsystems can lead to system-wide compromise, motivating runtime monitoring and layered defenses in consolidated architectures \cite{Experimental_security_analysis_of_a_modern_automobile}\cite{Remote_exploitation_of_an_unaltered_passenger_vehicle}.

\section{Threat model}
\label{sec:threatM}

In this section, we define the threat model for next generation vehicles, focusing on adversaries targeting the software stack,
particularly the POSIX-based operating systems and their components. Our model is structured around adversary goals, capabilities,
and the vehicle's attack surface, following established cybersecurity frameworks \cite{iso21434}\cite{kifor2024automotive}.

Focusing vulnerability analysis on the OS, middleware, virtualization, and application layers addresses the software stack where major threats emerge in next-generation vehicles—particularly given the integration of POSIX components in infotainment systems \cite{sayad2023cyber}\cite{jeong2023infotainment}  and enables faster vulnerability discovery and remediation by leveraging established POSIX-compatible security tools.

\subsection{Adversary goals}
An adversary's primary objectives can range from simple mischief to life-threatening attacks. We consider the following goals:
\begin{itemize}
    \item \textbf{Compromise vehicle safety:} interfere with safety-critical functions such as braking, steering, or acceleration. This is the most severe threat \cite{eiza2017cybersecurity}.
    \item \textbf{Data theft:} steal sensitive personal or vehicle data, including location history, contact information, or telemetry data that could reveal driving habits \cite{kifor2024automotive}.
    \item \textbf{Unauthorized access and control:} gain unauthorized control over vehicle functions, e.g., unlocking doors, starting the engine, or settings manipulation~\cite{uddin2023systematic}.
    \item \textbf{Financial gain:} deploy ransomware to disable the vehicle until a payment is made, steal subscription-based services \cite{huq2024cybersecurity} or, being the owner of the vehicle, bypass toll collection systems' payment \cite{lembhe2023toll}.
    \item \textbf{Large scale disruption:} execute a fleet-wide\footnote{A malware compromises large numbers of IoT devices or vehicles} attack, potentially causing widespread traffic disruption or reputational damage to the manufacturer \cite{he2018security}.
\end{itemize}

The automotive industry's most important problem is to ensure that fleet-wide attacks cannot take place, or at least with minimal consequences \cite{grimm2021contextaware}\cite{malik2020analysis}.
Such a scenario could lead to a whole fleet of vehicles being taken over remotely and used as a terrorist weapon or, for a mass theft of vehicles and even the sale of stolen user data.

\subsection{Adversary capabilities and access}
We classify adversaries based on their skill level, resources, and the type of access they have to the vehicle \cite{eiza2017cybersecurity}.

\begin{description}
    \item[Remote attacker:] this adversary has no physical access to the vehicle and attempts to compromise it over a wireless interface (e.g., cellular, Wi-Fi, Bluetooth). We distinguish between:
    \begin{itemize}
        \item \textit{Low sophistication:} uses publicly known vulnerabilities and exploit scripts. Their goal is often data theft or minor control functions \cite{uddin2023systematic}.
        \item \textit{High sophistication:} possesses the resources to discover zero-day vulnerabilities, holding deep understanding of automotive systems. Their goals can include large-scale disruption or persistent espionage \cite{kifor2024automotive}.
    \end{itemize}

    \item[Physical attacker:] this adversary has direct physical access to the vehicle's internal systems. This category includes:
    \begin{itemize}
        \item \textit{Malicious insider:} a  mechanic, technician, or employee with legitimate access who abuses their privileges \cite{nhtsa2022cybersecurity}.
        \item \textit{Technically skilled owner:} an owner who attempts to modify their vehicle's software, potentially introducing vulnerabilities unintentionally \cite{eiza2017cybersecurity}.
        \item \textit{Thief:} an attacker who gains temporary physical access to an unattended vehicle to connect to physical ports like the OBD-II or USB \cite{eiza2017cybersecurity}.
    \end{itemize}
\end{description}

In our work, we aim to address both physical attackers and remote attackers, as the shift to connected, POSIX-based systems makes this vector increasingly probable~\cite{uddin2023systematic}.

\subsection{Attack surface}
The attack surface of a next-generation vehicle is extensive and spans hardware and software layers \cite{Advancing_Security_in_Software_Defined_Vehicles_A_Comprehensive_Survey_and_Taxonomy}. We identify the following key areas \cite{kifor2024automotive}:

\begin{itemize}
    \item \textbf{Connectivity interfaces:} any component that communicates with the outside world is a potential entry point. This includes cellular (4G/5G) modems, Wi-Fi and Bluetooth radios, and V2X communication units~\cite{eiza2017cybersecurity}.
    \item \textbf{Infotainment system (IVI):} become complex systems, running third-party applications and web browsers, they represent a significant portion of the attack surface. A vulnerability in an application or the underlying OS (e.g., AGL or AAOS) can be an initial entry point\cite{huq2024cybersecurity}.
    \item \textbf{Over The Air (OTA) updates:} the mechanism for delivering software updates is a high-value target. A compromised OTA update could lead to a fleet-wide attack~\cite{he2018security}.
    \item \textbf{Physical ports:} exposed interfaces like the OBD-II diagnostic port and USB ports provide a direct line of access to the vehicle's internal networks~\cite{eiza2017cybersecurity}.
    \item \textbf{Supply chain:} vulnerabilities introduced via compromised third-party software libraries or hardware components from suppliers. The use of open-source software in POSIX-based systems makes this a critical concern~\cite{nhtsa2022cybersecurity}.
\end{itemize}


This threat model specifies the adversarial assumptions and attack surfaces considered, thereby guiding both the design choices of VERA and the criteria used to evaluate its effectiveness. 
By systematically identifying CVEs in the vehicle's software stack, our tool helps to mitigate these threats by enabling developers to find and patch vulnerabilities before they can be exploited \cite{kifor2024automotive}.

\section{The VERA suite}
\label{sec:vera}

\subsection{Architecture}

We present in this section our Vulnerability Exposure and Reporting Analysis (VERA) suite, a lightweight, offline-capable, and scalable scanner designed to detect CVEs across container images, Android devices/emulators and product/package inventories.

VERA is optimized for batch analyses (lists of images or products). 
The scanner integrated into the suite leverages a local CVE dataset: MITRE's database, CVE List V5\footnote{\url{https://github.com/CVEProject/cvelistV5}} and combines \texttt{pip-audit},\footnote{\url{https://pypi.org/project/pip-audit/}} the \texttt{Docker Software Development Kit},\footnote{\url{https://docs.docker.com/reference/api/engine/sdk/}} and \texttt{Dask}\footnote{\url{https://www.dask.org/}} for parallel execution to deliver fast, configurable, and reproducible results.

We designed VERA's experimental testbed using Docker containers because they provide an optimal environment for scalable, automated, and easily reproducible vulnerability scanning. While we acknowledge that isolated containers do not fully replicate a production hardware deployment, they perfectly preserve the crucial elements required to validate our semi-white-box approach: the exact filesystem, installed packages, and software configurations. Consequently, this environment ensures high fidelity for our core contributions (vulnerability detection, parsing, and filtering) while the impact of hardware-specific defenses on runtime exploitability remains for future work.

\begin{figure}[t!]
	\begin{center}
	\includegraphics[width=0.48\textwidth]{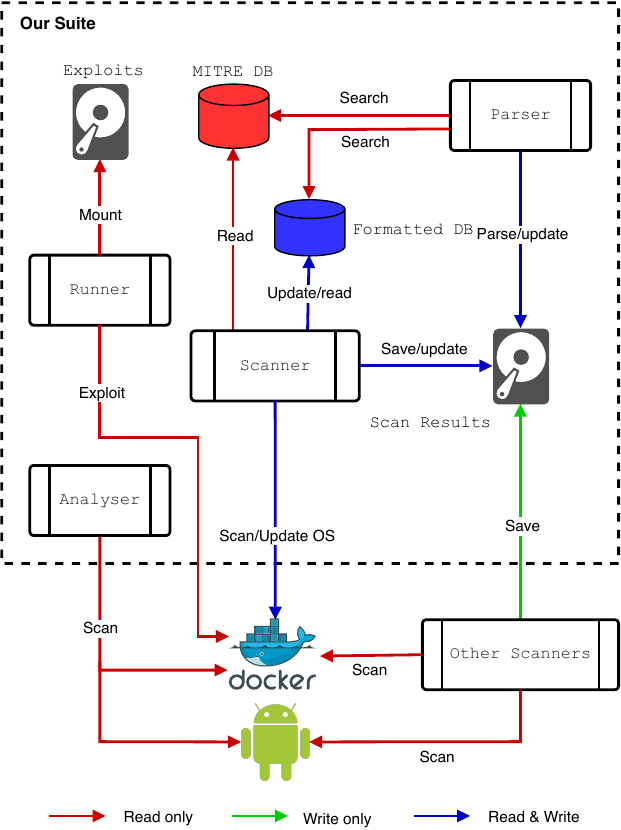}
	\caption{Architecture of VERA.}
	\label{fig:cvechecker_architecture}
	\end{center}
\end{figure}

Figure \ref{fig:cvechecker_architecture} describes VERA's architecture.
\texttt{Other Scanners} refers to tools like \texttt{Vanir}\footnote{\url{https://github.com/google/vanir}} or CVE Binary Tool (CBT)\footnote{\url{https://github.com/ossf/cve-bin-tool}}.

To use the built-in scanner \texttt{Scanner}, it has to
(1) create a new database \texttt{Formatted DB} in a custom standard format based on \texttt{MITRE DB}, to eventually
(2) scan a Docker image. It is also possible to use another scanner cf. \texttt{Other Scanners}, where CBT would allow to scan an Android emulator through Android Debug Bridge (ADB), Grype can be used to scan a Docker image, Vanir and Yocto's \texttt{cve-check}\footnote{\url{https://docs.yoctoproject.org/dev/dev-manual/vulnerabilities.html}} can be used to scan Android-based and AGL source code. These scans can be realised by another tool we made in order to automate eventual pulling from ADB, and ensure the right configurations are used for the scanners e.g output format, path to save.
(3) All the scan results are saved under a specific directory cf. \texttt{Scan Results} where
(4) they can be parsed by our tool \texttt{Parser}.
\texttt{Parser} allows to merge, filter and sort results, here is done the filtering to exclude unexploitable CVEs in the context of next-gen vehicles.
It also allows to see available exploits for each CVE,
(5) so \texttt{Analyser} can more particularly be used to ensure a vulnerability is present, searching a specific symbol, and see by which binaries it is called, and
(6) eventually download an exploit under a specific directory \texttt{Exploits} that can be later mounted in one Docker image through \texttt{Runner}.

\subsection{Methodology}

Referring back to Figure~\ref{fig:layers}, our framework applies to any OS, middleware, and application layer, and can also be used with POSIX‑compliant virtualization platforms such as QNX Neutrino or INTEGRITY. More precisely, our solution would scan binaries (including libraries) through Grype, CBT, or our built-in scanner, and for repo scanners through Vanir and \texttt{cve-check} for Android and Yocto-based OS.
Then, filtering removes package classes that would be meaningless in the context of next-generation vehicles, especially (1) non-sudo command-line utilities, or softwares that cannot lead to a privilege escalation, or remote code execution in an automotive setting (e.g apt, curl, git, ssh); or (2) developer toolchains and interpreters (e.g gcc, cmake, perl).
Indeed, we position ourselves as an attacker that have the same capabilities as on a genuine vehicle, where a victim does not have a Command Line Interface (CLI), excluding all CLI tools except sudo commands
as an attacker who gained access could exploit them for a later privilege escalation (LPE).
It still remains possible to view these filtered packages, for instance, an attacker might want to see GCC vulnerabilities in order to bypass Address Space Layout Randomization (ASLR).
Therefore, this filtering shall be seen more as a helper for potential vulnerabilities directly exploitable.

The proposed framework is comprehensive because beyond broad identification, it incorporates environment‑specific validation through a dockerized setup that mirrors realistic conditions, enabling reproducible and scalable testing. This combination of vulnerability coverage, prioritization, exploitability evaluation, and cross‑stack applicability allows the framework to reveal both the breadth of potential threats and the practical constraints that shape real‑world attack feasibility. By integrating these components into a single workflow, the framework delivers a holistic view of SDV security exposure that existing tools --- typically limited to scanning --- cannot achieve.

Figure~\ref{fig:cvechecker_workflow} describes the main workflows associated to VERA. For simplicity, the figure omits configuration options and OS image updates.
Each updatable OS image, including its packages, was upgraded to the latest available versions and then scanned, as summarized in Table~\ref{tab:cve-cvss-benchmark}. Android and Yocto‑based images could not be upgraded and were therefore analyzed in their original form (Table~\ref{tab:cve-cvss-benchmark-2}).
The processing pipeline is as follows:

\begin{figure*}[t!]
	\centering
	\subfloat[]{\includegraphics[width=0.42\textwidth]{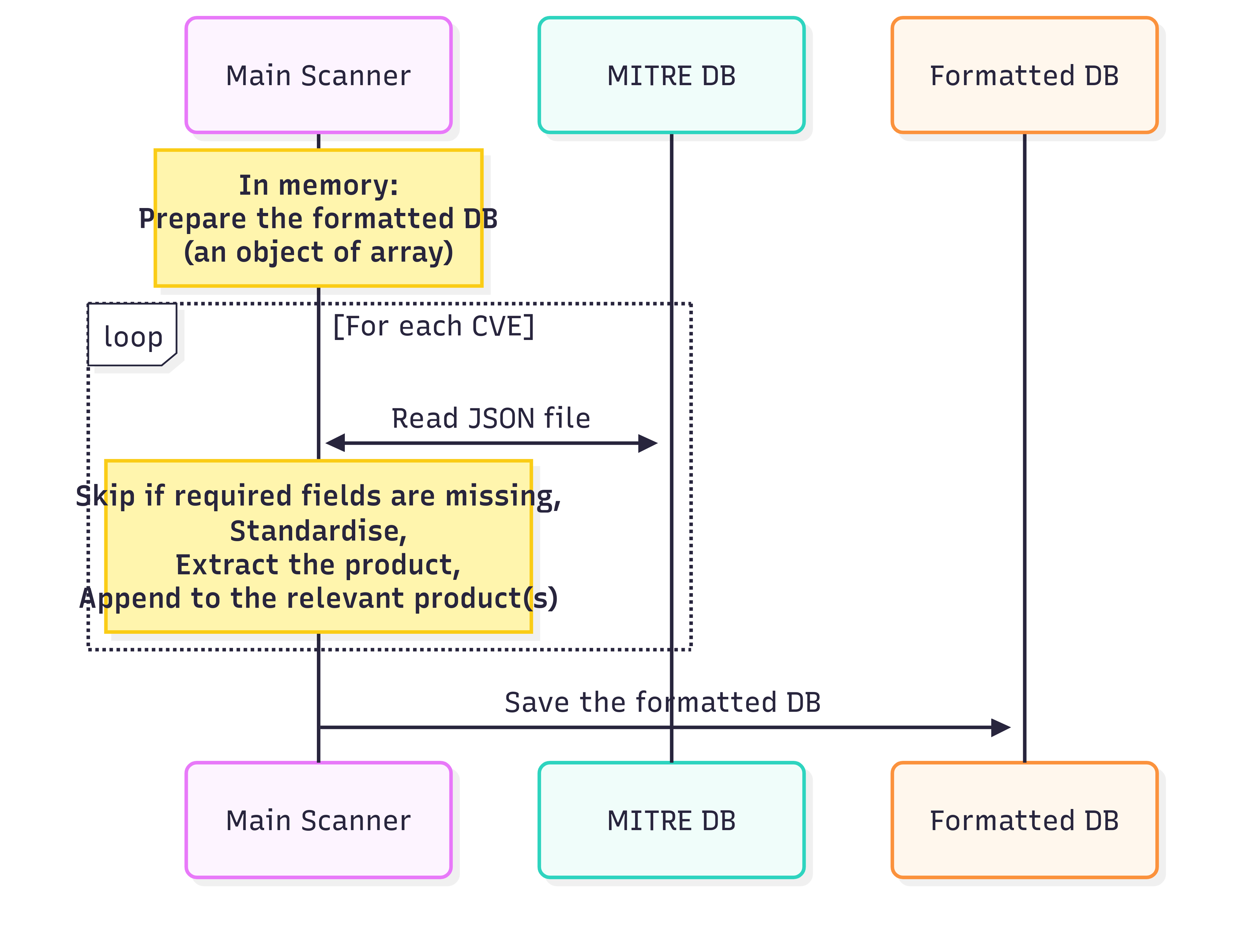}}
	\hspace{0.2cm}
	\subfloat[]{\includegraphics[width=0.48\textwidth]{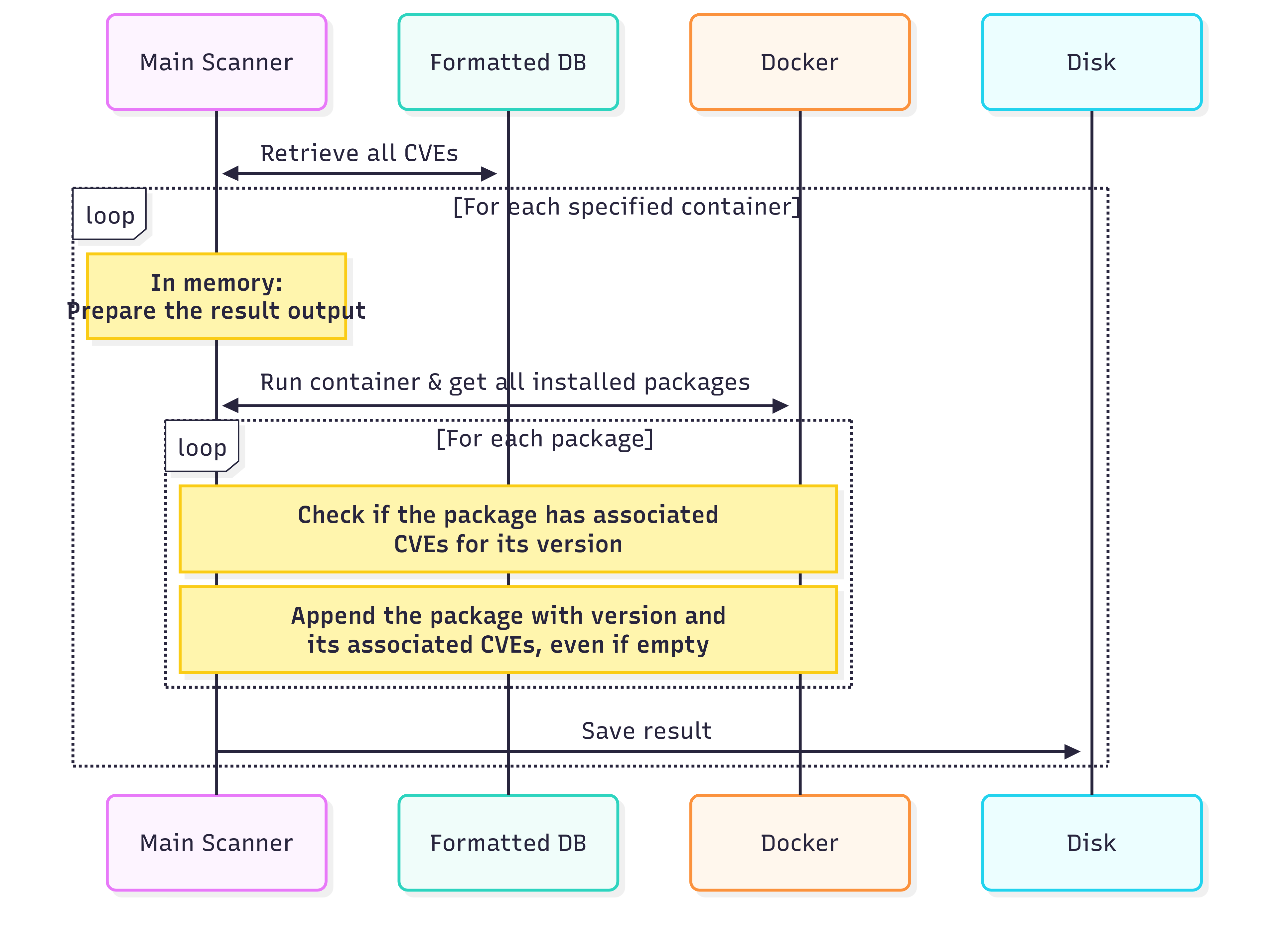}}
	\\
	\subfloat[]{\includegraphics[width=0.82\textwidth]{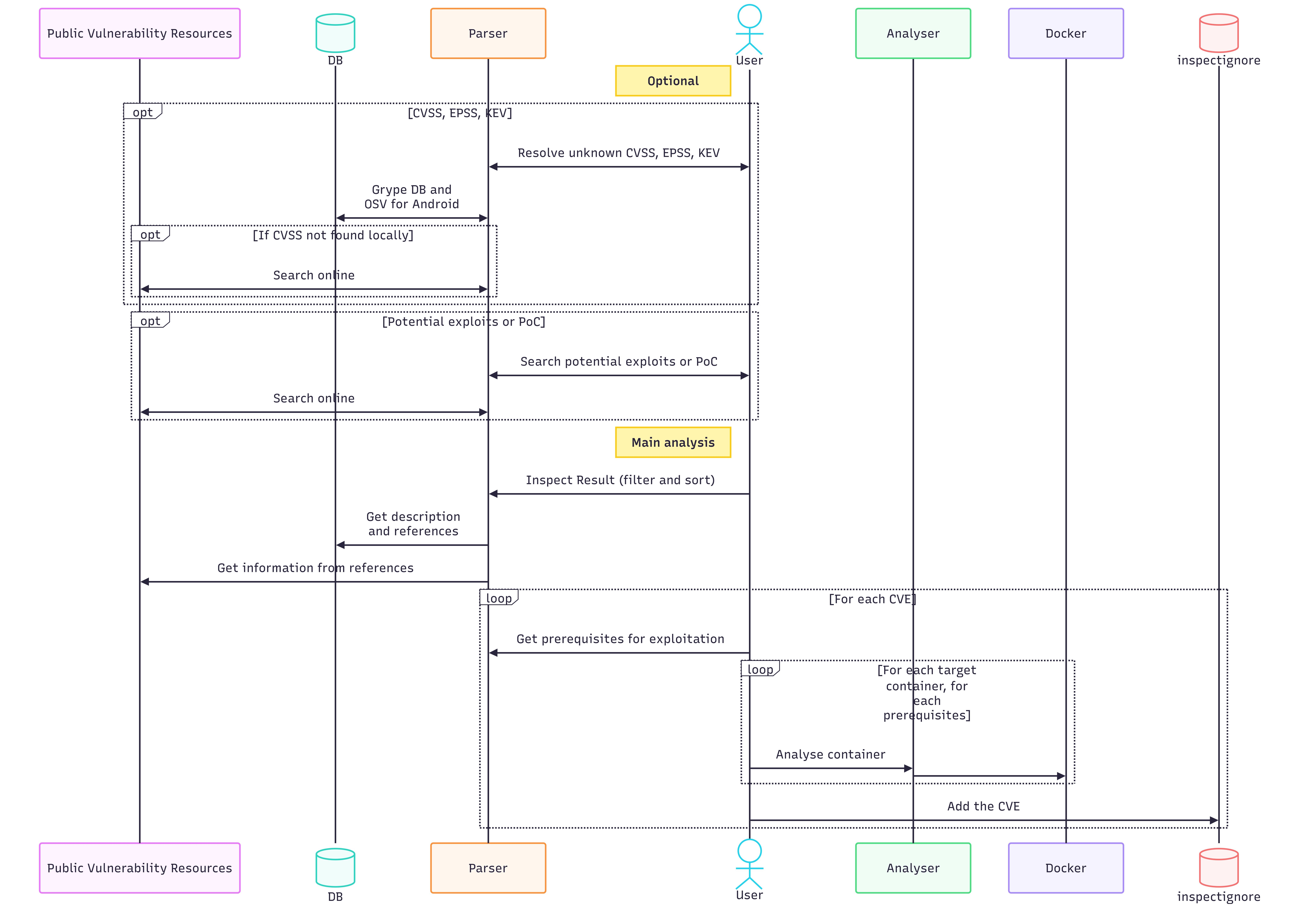}}
	\caption{Workflow of VERA: (a) setup, (b) Docker images/Android devices scanning, (c) Results analyses  and exploitability assessment. 
	DB: other scanners vulnerability databases (e.g., Grype, OSV)}
	\label{fig:cvechecker_workflow}
\end{figure*}
\begin{enumerate}
  \item \textbf{Inputs:} accepts Docker images, package lists or package files as the analysis seed.
  \item \textbf{Normalization and preprocessing:} parses, minify and normalize package/version traces. More precisely, helper scripts extract package lists from images or host outputs and convert them to the canonical format.
  \item \textbf{Systematic evaluation and scoring:} For each detected issue, the pipeline attaches canonical severity metadata when available (e.g., CVSS vector/score from the CVE record). If no CVSS is present, VERA queries authoritative sources (e.g. NIST and vendor advisories such as Red Hat). 
  \item \textbf{Results generation:} outputs JSON structured results containing matched CVEs, their scores CVSS and EPSS.
  \item \textbf{Inspection and reporting:} provides lightweight inspection utilities (e.g., \texttt{parser.sh inspect}, \texttt{parser.sh table}) for summaries, with the possibility to filter by severity, ecosystem, package name or type, and exclude CVEs once reviewed (based on the directory \texttt{inspectignore}).
\item \textbf{Assess Exploitability:} VERA can analyse a container OS (e.g., symbols, changelogs, SELinux configuration) to assess CVEs exploitability and identify false positives.
\end{enumerate}

\begin{table}[!ht]
    \centering
    \caption{Simple package filtering. \textbf{Pre:} No filtering, \textbf{Flt:} Filtering applied, \textbf{Diff:} Difference between raw result and filtered result.}
    \label{tab:benchmark-filtering}
    \begin{tabular}{lll|ll}
    \toprule
        \textbf{OS} & \textbf{Pre} & \textbf{Flt} & \multicolumn{2}{l}{\textbf{Diff}} \\ \midrule
        Eclipse S-CORE & 9 & 8 & 1 & 11.1\% \\
        Ubuntu 22.04 & 22 & 13 & 9 & 40.9\% \\
        Ubuntu 20.04 & 22 & 13 & 9 & 40.9\% \\
        VxWorks 7 & 77 & 33 & 44 & 57.1\% \\
        VxWorks 7 ROS2 & 100 & 51 & 49 & 49.0\% \\
        QNX Neutrino & 100 & 56 & 44 & 44.0\% \\
        Zephyr & 193 & 91 & 102 & 52.8\% \\
        TeslaOS & 280 & 243 & 37 & 13.2\% \\
        AutoSD & 321 & 129 & 192 & 59.8\% \\
        ROS2 & 331 & 281 & 50 & 15.1\% \\
        AAOS 34 & 160 & 159 & 1 & 0.6\% \\
        Android 32 & 634 & 627 & 7 & 1.1\% \\
        Android 30 & 757 & 749 & 8 & 1.1\% \\
        AGL & 1217 & 1203 & 14 & 1.2\% \\
        \bottomrule
    \end{tabular}
\end{table}

\begin{table}[!ht]
    \centering
    \caption{CVEs classes by OS. \textbf{RCE:} Remote Code Execution, \textbf{LPE:} Local Privilege Escalation, \textbf{ID:} Information Disclosure (cause data leakage), \textbf{DoS:} Denial of Service (cause disruption).}
    \label{tab:benchmark-vuln-type}
    \begin{tabular}{llllll}
        \toprule
        \textbf{OS} & \textbf{RCE} & \textbf{LPE} & \textbf{ID} & \textbf{DoS} & \textbf{Unknown} \\ \midrule
        Eclipse S-CORE & 2 & 0 & 1 & 3 & 2 \\
        Ubuntu 22.04 & 4 & 1 & 3 & 3 & 2 \\
        Ubuntu 20.04 & 4 & 1 & 3 & 3 & 2 \\
        VxWorks 7 & 7 & 4 & 2 & 11 & 9 \\
        VxWorks 7 ROS2 & 8 & 7 & 5 & 17 & 14 \\
        QNX Neutrino & 10 & 6 & 3 & 19 & 18 \\
        Zephyr & 10 & 8 & 8 & 42 & 23 \\
        TeslaOS & 20 & 47 & 63 & 85 & 28 \\
        AutoSD & 29 & 2 & 30 & 42 & 26 \\
        ROS2 & 70 & 34 & 28 & 94 & 55 \\
        
        AAOS 34 & 14 & 81 & 25 & 34 & 5 \\
        Android 32 & 63 & 307 & 148 & 97 & 12 \\
        Android 30 & 87 & 337 & 202 & 102 & 21 \\
        AGL & 226 & 110 & 77 & 614 & 176 \\
        \bottomrule
    \end{tabular}
\end{table}

Our solution yields several practical advantages: (1) VERA is not just a scanner, but a suite of tools, that filters, sorts and prioritizes CVEs; (2) It provides functionality to scan layered/Open Container Initiative (OCI) container images or Android emulator directly; (3) it provides guidance to assess exploitability, in addition to discovering online exploits.

\subsection{Complementary scanners}

\begin{table*}[!ht]
	\centering
	\scriptsize
    \setlist[itemize,1]{leftmargin=*,labelindent=0mm,labelsep=2.0mm,itemsep=0.6mm}
    \setlength{\extrarowheight}{2pt}

	\begin{tabular}{l|p{7.6cm}|p{7.2cm}}
		\textbf{Scanner} & \textbf{Pros} & \textbf{Cons} \\ \hline
		\textbf{OSV Scanner} & \begin{itemize}
			\item   Accurate open-source dependency scanning via OSV.dev
			\item   Container scanning
			\item   Tracks vulnerability states (e.g., fixed, wont-fix).
			\end{itemize} & \begin{itemize}
			\item   No EPSS/risk scoring,
			\item   Cannot scan large containers
			\item   Seems less efficient on non-Debian-based OSes
			\end{itemize} \\ \hline
		\textbf{CVE Binary Tool} &
			\begin{itemize}
			\item   Scans compiled binaries for known vulnerable libs (e.g OpenSSL, zlib)
			\item   Works offline, simple setup
			\item   Extract \& Scan archives (e.g zip, tar), including APKs
            \item   EPSS and KEV Indicator
			\end{itemize} &
			\begin{itemize}
			\item   No container scanning
			\item   No state tracking
			\item   Prone to false negatives/positives due to string matching in binary
			\item   Incomplete Android support
			\item   Much slower than other scanners
			\end{itemize} \\ \hline
		\textbf{Trivy} &
			\begin{itemize}
			\item   Full-featured: containers, filesystems, IaC, dependencies, secrets
			\item   Supports risk scoring (CVSS, EPSS)
			\item   Tracks vulnerability states (fixed/unfixed)
			\end{itemize} &
			\begin{itemize}
			\item   Heavier runtime and slower on large images
			\item   Seems more prone to false positives/negatives 
			\item   No KEV Indicator
			\end{itemize} \\ \hline
		\textbf{Grype} &
			\begin{itemize}
			\item   Container and SBOM-based scanning (via Syft)
			\item   Tracks vulnerability states (fixed/unfixed)
			\item   Supports risk scoring (CVSS, EPSS) and KEV Indicator
			\end{itemize} &
			\begin{itemize}
			\item   Slightly slower than Trivy
			\end{itemize} \\ \hline
        \textbf{VERA (Ours)} &
        \begin{itemize}
            \item   Upgrade dependencies before scanning container
            \item   Android device/emulator scanning (via CBT)
            \item   Resolve unknown CVSS and EPSS
            \item   Standardize/compress reports for all scanners
            \item   Filter and sort results/reports
            \item   Fast checking CVE presence and exploitability
            \item   Search for CVE information and online exploits
        \end{itemize} &
			\begin{itemize}
			\item   No state tracking
			\item   Not full support of scanner reports (i.e vulnerability states, EPSS, KEV indicator)
			\end{itemize} \\ \hline
	\end{tabular}
	\caption{Scanners comparison}
	\label{tab:scanner_comparison}
\end{table*}

\begin{table}[!ht]
	\centering
	\scriptsize
    \setlist[itemize,1]{leftmargin=*,labelindent=0.4mm,labelsep=1.8mm,itemsep=0.8mm}
    
	\begin{tabular}{p{1.7cm}|l|p{4.3cm}}
		\textbf{Scanner} & \textbf{Target} & \textbf{Cons} \\ \hline
		\textbf{Vanir} & Android-based & \begin{itemize}
			\item   Do not show affected product's name/version
			\item   No CVSS/EPSS or KEV indicator
		\end{itemize} \\ \hline
		\textbf{Yocto's built-in \texttt{cve-check} ft.} & Yocto-based & \begin{itemize}
			\item   No EPSS or KEV indicator
		\end{itemize} \\ \hline
	\end{tabular}
	\caption{Repo scanners}
	\label{tab:repo_scanners}
\end{table}


Table \ref{tab:scanner_comparison} summarizes the pros and cons of the scanners the most used by the community, and compare them to our tool VERA.
Table \ref{tab:repo_scanners} summarizes the repo scanners we used for Android and Yocto-based OSes. 

All tracking-based scanners (e.g., Grype\footnote{\url{https://github.com/anchore/grype}}, OSV\footnote{\url{hhttps://github.com/google/osv-scanner}} or Trivy\footnote{\url{https://trivy.dev/}}), which rely on conventional vulnerability databases, are primarily designed to analyze package-based systems and built operating system images for which component metadata and versioning information are available through standard package databases or management mechanisms (such as \texttt{apt}, \texttt{rpm}, or \texttt{dpkg}) or via Software Bills of Materials (SBOMs). These tools perform vulnerability attribution by matching discovered package names and versions against curated vulnerability feeds. However, they cannot be directly applied to Android and Yocto-based operating systems using their default tracking mechanisms,
as these platforms rely on fundamentally different build architectures and deployment models that do not preserve standard package database in the final system image. As a result, vulnerability assessment for Android and Yocto typically requires analysis at the source level rather than at the level of the built image. In this context, Vanir is a vulnerability scanner developed by Google specifically for Android-based systems, which operates by matching source code and patch signatures and is designed to achieve low false-positive rates. Similarly, the Yocto Project provides a built-in vulnerability analysis mechanism, \texttt{cve-check} (referred to as \texttt{cve-check} in the remainder of this paper), which inspects Yocto recipes and source repositories to identify known vulnerabilities during the build process. Unlike scanners that operate on built OS images, both Vanir and \texttt{cve-check} rely on source-level tracking and platform-specific metadata, making them better suited for vulnerability detection on their respective operating systems.

This source-based approach, however, comes at the cost of increased setup complexity and longer analysis times: several hours may be required to download source trees and complete partial or full builds before scanning can be performed, in contrast to package-based scanners such as Grype or Trivy, which can typically analyze a built OS image or container within minutes. CVE Binary Tool (CBT) represents another example of a scanner applicable to Android-based and Yocto-based components, scanning binaries to retrieve their name and version.

Because our scanner is not designed to natively analyze Android-based systems, VERA leverages Vanir and CBT for Android-based OSes, and \texttt{cve-check} and CBT for Yocto-based OSes. 
Since Vanir and \texttt{cve-check} do not provide complete coverage of all CVEs in practice, we use CBT as a complementary scanner. CBT findings are considered unless the corresponding CVEs are identified as patched by Vanir or \texttt{cve-check}. 


\begin{table*}[!ht]
    \caption{CVE \& CVSS benchmark by OS.}
    	\setlength{\extrarowheight}{2pt}
	
    \label{tab:cve-cvss-benchmark}
	\begin{tabular}{l|rlrlrlrlrl|rl|rl|r|r}
		~ & \multicolumn{2}{c}{\textbf{\scriptsize{UNKNOWN}}}	&
			\multicolumn{2}{c}{\textbf{\scriptsize{LOW}}}	& \multicolumn{2}{c}{\textbf{\scriptsize{MEDIUM}}}	& \multicolumn{2}{c}{\textbf{\scriptsize{HIGH}}}	& \multicolumn{2}{c|}{\textbf{\scriptsize{CRITICAL}}}	&
			\multicolumn{2}{c|}{\textbf{\scriptsize{TOTAL}}}		& \multicolumn{2}{c|}{\textbf{\scriptsize{EPSS}}} &
			\textbf{KEV}		& \textbf{Exploits}
			\\
		\textbf{OS / Middleware}	&
			Gr & VE		& Gr & VE	& Gr & VE		& Gr & VE		& Gr & VE		&
			Gr & VE		& Gr & VE	&
			VE & VE \\ \hline
Eclipse S-CORE& 0   & 0    & 2       & 2    & 6          & 5    & 1        & 1    & 0  & 0    & 9         & 8    & 0        & 8   & 0 & 3   \\ \hline
Ubuntu 22.04  & 0   & 0    & 4       & 2    & 11         & 5    & 6        & 5    & 1  & 1    & 22        & 13   & 22       & 13  & 0 & 8   \\ \hline
Ubuntu 20.04  & 0   & 0    & 4       & 2    & 11         & 5    & 6        & 5    & 1  & 1    & 22        & 13   & 22       & 13  & 0 & 8   \\ \hline
VxWorks 7     & 0   & 0    & 18      & 7    & 39         & 13   & 19       & 12   & 1  & 1    & 77        & 33   & 77       & 33  & 0 & 20  \\ \hline
VxWorks 7 ROS2& 0   & 0    & 24      & 11   & 47         & 20   & 28       & 19   & 1  & 1    & 100       & 51   & 100      & 51  & 0 & 29  \\ \hline
QNX Neutrino  & 0   & 0    & 30      & 13   & 43         & 25   & 26       & 17   & 1  & 1    & 100       & 56   & 85       & 49  & 0 & 36  \\ \hline
Zephyr        & 0   & 0    & 34      & 14   & 90         & 40   & 62       & 36   & 7  & 1    & 193       & 91   & 190      & 90  & 0 & 52  \\ \hline
TeslaOS       & 0   & 0    & 28      & 13   & 186        & 169  & 65       & 60   & 1  & 1    & 280       & 243  & 44       & 240 & 2 & 119 \\ \hline
AutoSD        & 0   & 0    & 25      & 12   & 166        & 80   & 127      & 35   & 3  & 2    & 321       & 129  & 319      & 128 & 0 & 75  \\ \hline
ROS2          & 0   & 0    & 43      & 26   & 136        & 111  & 122      & 116  & 30 & 28   & 331       & 281  & 327      & 278 & 0 & 215 \\ \hline
\end{tabular}
	\centering
	\parbox{\linewidth}{
		\scriptsize
		\textbf{Legend:} \\
		Exploits: CVEs associated with publicly available proof-of-concept code or reported online exploits (non-exhaustive) \\
		CB: CBT, Gr: Grype, VC: Vanir (Android-based) or \texttt{cve-check} (Yocto-based), VE: VERA
	}

\vspace{0.4cm}

	\setlength{\tabcolsep}{4pt}
	\setlength{\extrarowheight}{2pt}
	
    \caption{CVE \& CVSS benchmark by OS. }
    \label{tab:cve-cvss-benchmark-2}
	\begin{tabular}{l|rrlrrlrrlrrlrrl|rrl|rrl|r|r}
		~ & \multicolumn{3}{c}{\textbf{\scriptsize{UNKNOWN}}}	&
			\multicolumn{3}{c}{\textbf{\scriptsize{LOW}}}	& \multicolumn{3}{c}{\textbf{\scriptsize{MEDIUM}}}	& \multicolumn{3}{c}{\textbf{\scriptsize{HIGH}}}	& \multicolumn{3}{c|}{\textbf{\scriptsize{CRITICAL}}}	&
			\multicolumn{3}{c|}{\textbf{\scriptsize{TOTAL}}}		& \multicolumn{3}{c|}{\textbf{\scriptsize{EPSS}}} &
			\textbf{KEV}		& \textbf{Exploits}
			\\
		\textbf{OS }	&
			\tiny{VC} & \tiny{CB} & \tiny{VE}		& \tiny{VC} & \tiny{CB} & \tiny{VE}		& \tiny{VC} & \tiny{CB} & \tiny{VE}		& \tiny{VC} & \tiny{CB} & \tiny{VE}		& \tiny{VC} & \tiny{CB} & \tiny{VE}		&
			\tiny{VC} & \tiny{CB} & \tiny{VE}		& \tiny{VC} & \tiny{CB} & \tiny{VE} &
			\tiny{VE}		& \tiny{VE}

			\\ \hline

        AAOS 34 &
			115 & 0 & 0 &
			0 & 5 & 5 &
			0 & 13 & 14 &
			0 & 19 & 129 &
			0 & 9 & 11 &
			115 & 46 & 159 &
			0 & 46 & 118 &
			3 & 24 \\ \hline

        Android 32 &
			547 & 1 & 1 &
			0 & 2 & 2 &
			0 & 31 & 71 &
			0 & 45 & 507 &
			0 & 17 & 53 &
			547 & 96 & 627 &
			0 & 96 & 591 &
			8 & 109 \\ \hline
		
        Android 30 &
			641 & 0 & 0 &
			0 & 4 & 4 &
			0 & 46 & 122 &
			0 & 53 & 560 &
			0 & 21 & 71 &
			641 & 124 & 749 &
			0 & 124 & 727 &
			4 & 102 \\ \hline

        AGL &
			0 & 1 & 1 &
			7 & 23 & 26 &
			437 & 463 & 844 &
			130 & 213 & 292 &
			6 & 51 & 40 &
			580 & 751 & 1203 &
			576 & 749 & 1201 &
			3 & 710 \\ \hline
	\end{tabular}
\end{table*}

\begin{table}[!ht]
    \caption{
        Overall comparison of total CVEs/EUVDs for each OS, including their type, purpose, and their certifications ISO/IEC 15408-5 (EAL) and ISO 26262 (ASIL).
	}
    \label{tab:cves-comparison}
	\newcolumntype{S}{>{\scriptsize}c}
	\newcolumntype{L}{>{\scriptsize}l}
    \centering
    \small
    \setlength{\extrarowheight}{2pt}
	\color{black}
    
    \begin{tabular}{lSSSSc}
        \textbf{OS / Middleware} & \textbf{Type} & \textbf{Purpose} & \textbf{EAL} & \textbf{ASIL} & \textbf{CVEs} \\ 
        \hline

        Eclipse S-CORE & \emptycircle & \filledcircle & \none & \none & 8 \\ \hline
        Ubuntu 22.04 & \multirow{2}{*}{\filledcircle} & & 2+ & \none & 13 \\
        Ubuntu 20.04 &  & & 2+ & \none & 13 \\ \hline
        VxWorks 7 & \multirow{2}{*}{\filledcircle} & \multirow{2}{*}{\filledcircle} & 6+ & D & 33 \\
        VxWorks 7 ROS2 & & & & & 51 \\ \hline
        QNX Neutrino & \filledcircle & \filledcircle & 4+ & D & 56 \\ \hline
        Zephyr & \filledcircle & \filledcircle & \none & \none & 91 \\ \hline
        AutoSD / RHIVOS & \filledcircle & \filledcircle & \none & B & 129 \\ \hline
		AAOS 34 & \filledcircle & \emptycircle & \none & \none & 159 \\ \hline
        TeslaOS & \filledcircle & \emptycircle & \none & \none & 243 \\ \hline
        ROS2 & \emptycircle & \emptycircle & \none & \none & 281 \\ \hline
		Android 32 & \multirow{2}{*}{\emptycircle} & \multirow{2}{*}{\emptycircle} & \none & \none & 627 \\
		Android 30 &  &  & \none & \none & 749 \\ \hline
        AGL & \filledcircle & \emptycircle & \none & \none & 1203 \\ \hline

		INTEGRITY	& \filledcircle & \filledcircle & 6+ & D & N/A \\ \hline
		PikeOS		& \filledcircle & \filledcircle & 5+ & D & N/A \\ \hline
		DRIVE OS 6	& \multirow{2}{*}{\filledcircle} & \multirow{2}{*}{\filledcircle} & \none & D & N/A \\
		DRIVE OS 5	&  &  & \none & B & N/A \\ \hline
		LynxOS      & \filledcircle & \filledcircle & \none & D & N/A \\ \hline
		Mentor Nucleus  & \filledcircle & \filledcircle & \none & D & N/A \\ \hline
    \end{tabular}
	\parbox{\linewidth}{
		\scriptsize
		~~\\
		\textbf{Legend:} \\
		Type:
		\filledcircle\ = OS, 
		\emptycircle\ = Middleware.\\
		Purpose:
		\filledcircle\ = Safety-critical, 
		\emptycircle\ = Non-safety-critical. \\
		CVE: N/A = non available for scanning
	}
\end{table}

\begin{table}[!ht]
    \centering
    \small
    \color{black}
    \caption{Tested OS details.}
    \label{tab:os-details}
    \begin{tabular}{llll}
        OS & Version & Release & Underlying OS \\ \hline
        AAOS & SDK 34 & & ~ \\ \hline
        Android & SDK 32 & & ~ \\ \hline
         & SDK 30 & & ~ \\ \hline
        ROS2 & ~ & humble & Ubuntu 22.04 \\ \hline
        VxWorks & 7 & humble & Ubuntu 22.04 \\ \hline
        ~ & 7 ROS2 & humble & Ubuntu 22.04 \\ \hline
        TeslaOS & amd-5.4.265 & focal & Ubuntu 20.04 \\ \hline
        QNX Neutrino & 8.0 & focal & Ubuntu 20.04 \\ \hline
        AGL IVI Demo Qt & 20.0.1 & trout &  \\ \hline
        AutoSD & 9 & / & CentOS 9 \\ \hline
        Eclipse S-CORE & 158 & / & Alpine 3.21.5 \\ \hline
        Zephyr & 0.28.7 & / & Ubuntu 24.04 \\ \hline
    \end{tabular}
\end{table}

\begin{table*}[!ht]
	\centering
	\scriptsize
       \caption{Examples of attack scenarios.}
       \label{tab:attack_scenario}
	\begin{tabular}{p{0.5cm}|p{1.4cm}p{4.0cm}|p{4.8cm}|cccc}
		&&&& \multicolumn{4}{c}{\textbf{Impact (Negligible --$>$ Severe)}} \\
		\textbf{Path} & \textbf{Tactic} & \textbf{Technique} & \textbf{Description} & Safety & Financ. & Operat. & Privacy \\
		\hline
		A & Initial Access & ATM-T0014: Malicious App &  Install a backdoor in the isolated execution environment (e.g., AAOS sandbox) via a malicious application from the infotainment's app store &
			\multirow{8}{*}{Severe} & \multirow{8}{*}{Major} & \multirow{8}{*}{Moderate} & \multirow{8}{*}{Negligible} \\

		~ & Affect Vehicle Function & ATM-T0071: Unintended Vehicle Network Message & Inject spoofed CAN/AE messages (e.g, wheel-speed or torque-request messages) targeting engine/transmission ECUs to cause acceleration or braking \\
		~ & ~ & ~ & ~ & ~ \\\hline

		B & Initial Access & ATM-T0011: Browser Compromise & Compromise the infotainment system via any disclosed vulnerability & 
			\multirow{12}{*}{Moderate} & \multirow{12}{*}{Moderate} & \multirow{12}{*}{Moderate} & \multirow{12}{*}{Severe} \\
		~ & Persistence & ATM-T0023: Modify Isolated Execution Environment & Ensure continued access across reboots/updates by implanting a backdoor in the isolated execution environment \\
		~ & Collection & ATM-T0058: Capture Camera or Audio & \multirow{3}{*}{\shortstack[l]{Collect various private data (conversations, SMS,\\ location etc.)}} \\
		~ & ~ & ATM-T0035: Capture SMS Message & \\
		~ & ~ & ATM-T0043: Location Tracking & \\
		~ & Exfiltration & ATM-T0063: Internet Communication & Compromised ECU's internet connection to exfiltrate data \\
		~ & ~ & ~ & ~ & ~ \\\hline
	\end{tabular}
\end{table*}

\begin{table}
    \caption{Feasibility factor values}
    \label{tab:feasibility_factors}
	\centering
	\scriptsize
	\begin{tabular}{rl|rl|rl}
		\textbf{Elapsed Time}             & \textbf{Value} & \textbf{Expertise}          & \textbf{Value}  & \textbf{Knowledge}        & \textbf{Value}  \\
		\hline
		$\leq$ 1 day             & 0 & Layman                 & 0 & Public           & 0 \\
		$\leq$ 1 week            & 1 & Proficient             & 3 & Restricted       & 3 \\
		$\leq$ 1 month           & 4 & Expert                 & 6 & Sensitive        & 7 \\
		$\leq$ 3 months          & 10 & Mult. experts         & 8 & Critical         & 11 \\
		$\leq$ 6 months          & 17 &                       &   &          &  \\
		$>$ 6 months             & 19 &                       &   &          &  \\
		\\
		\textbf{Opportunity}               & \textbf{Value} & \textbf{Resources} & \textbf{Value}        \\
		\hline
		Unlimited & 0 & Standard         & 0 \\
		Easy      & 1 & Specialized      & 4 \\
		Moderate  & 4 & Bespoke          & 7 \\
		Difficult & 10 & Mult. bespoke   & 9 \\
	\end{tabular}
\end{table}

\begin{table}[!ht]
    \caption{Attack Potential/Feasibility qualitative scale}
    \label{tab:feasibility_scale}
	\centering
	\scriptsize
	\begin{tabular}{r|ll}
		\textbf{Attack Potential} & \textbf{Feasibility} \\ \hline 
		\textbf{0 -- 9}           & High		\\
		\textbf{10 -- 19}         & Medium		\\
		\textbf{20 -- 29}         & Low			\\
		\textbf{$\geq$ 30}        & Very Low	\\
	\end{tabular}
\end{table}

\begin{table*}[!ht]
	\newcolumntype{T}{>{\tiny}c}
	\centering
	\scriptsize
       \caption{Feasibility and risk of attack scenarios' example.}
       \label{tab:attack_feasibility}
	\begin{tabular}{c|p{1.6cm} >{\tiny}p{2.8cm}|TTTTT|c|c}
		\textbf{Path} & \textbf{Tactic} & \textbf{\scriptsize{Technique}}	& \textbf{Elapsed Time} & \textbf{Expertise} & \textbf{Knowledge} & \textbf{Opportunity} & \textbf{Resources} & \textbf{Feasibility} & \textbf{Vulnerabiliy ex.} \\
		\hline
		\textbf{A} & Initial Access & ATM-T0014: Malicious App				& $\leq$ 1 week & Proficient & Public & Easy & Standard & High & \tiny{CWE-912: Hidden Functionality} \\
		~ & Affect Vehicle Function & ATM-T0071: Unintended Vehicle Network Message
			& $\leq$ 6 months & Expert & Public & Unlimited & Standard & Low & De-association \cite{daniel2021someip} \\
		\cline{2-9}
		 ~ & \multicolumn{7}{r}{\textbf{Overall Feasibility:}} & Low \\
		\\ \hline

		\textbf{B} & Initial Access & ATM-T0011: Browser Compromise			& $\leq$ 6 months & Mult. Experts & Public & Easy & Standard & Low & CVE-2022-35737 \\
		~ & Persistence & ATM-T0023: Modify Isolated Execution Environment	& $>$ 6 months & Mult. Experts & Public & Unlimited & Standard & Low & \tiny{CWE-284: Improper Access Control} \\
		~ & Collection & ATM-T0058: Capture Camera or Audio					& $\leq$ 1 week & Proficient & Public & Unlimited & Standard & High & \na \\
		~ & ~ & ATM-T0035: Capture SMS Message & ~ & ~ & ~ & ~ & ~ & \\
		~ & ~ & ATM-T0043: Location Tracking & ~ & ~ & ~ & ~ & ~ & \\
		~ & Exfiltration & ATM-T0063: Internet Communication				& $\leq$ 1 day & Layman & Public & Unlimited & Standard & High & \na \\
		\cline{2-9}
		 ~ & \multicolumn{7}{r}{\textbf{Overall Feasibility:}} & Low \\
		\\ \hline

	\end{tabular}
	\parbox{\linewidth}{\scriptsize
	~~\\
	\textbf{Legend:}
	\na: Not applicable: weakness is due to core design.
	}
\end{table*}

\begin{table}[!ht]
    \caption{Risk Matrix.}
    \label{tab:risk_matrix}
	\centering
	\scriptsize
	\begin{tabular}{l|llll}
		\textbf{Impact} $\rightarrow$ \\
		\textbf{Feasibility} $\downarrow$ & Negligible & Moderate & Major & Severe \\
		\hline
        High		& Very Low	& Medium	& High		& Very High \\
        Medium		& Very Low	& Low		& High		& High		\\
        Low			& Very Low	& Low		& Medium	& Medium 	\\
        Very Low	& Very Low	& Very Low	& Low		& Low		\\
	\end{tabular}
\end{table}

\begin{table}
    \caption{Attack scenarios weighted risk.}
    \label{tab:attack_scenario_risks}
	\centering
	\scriptsize
	\begin{tabular}{r|cccc|c}
		& \multicolumn{4}{c}{\textbf{Risk}} \\
		\textbf{Path} & Safety & Financ. & Operat. & Privacy & Overall \\
		\hline
		A & Medium	& Medium	& Low	& Very Low	& Medium \\
		B & Low		& Low		& Low	& Medium	& Medium \\
	\end{tabular}
\end{table}

\section{Experimental evaluation and use case analysis}
\label{sec:experimental}

We present in this section an experimental evaluation over two representative use cases, designed to assess the practical impact of some identified vulnerabilities (i.e., to evaluate the full exploitation workflow, from CVE detection to end-to-end exploitability, thereby validating the relevance of the scan results under realistic attack conditions). As for the quantitative vulnerability analysis, Table \ref{tab:cve-cvss-benchmark} reports the CVE and CVSS benchmarks for each operating system, together with the subset of vulnerabilities classified as Known Exploited Vulnerabilities (KEV), that is, CVEs that are actively exploited in the wild
\textcolor{black}{; with potential online exploits -- only for CVEs with CVSS superior or equal to 5.0, and EPSS superior or equal to 0.0003 (0.03\%)}.
Table \ref{tab:cve-cvss-benchmark-2} is specific for Android and Yocto-based OSes, due to their different scanning methods (whose results are based on Vanir, \texttt{cve-check} merged with CBT and filtered by their relevance). Note that the default filtering configuration used in our experiments, as reported in Tables~\ref{tab:cve-cvss-benchmark}, \ref{tab:cve-cvss-benchmark-2}, \ref{tab:cves-comparison} is limited to package-level filtering. In other words, it does not exclude any broad vulnerability class by default. The vulnerability classes discussed in the analysis (e.g., RCE, LPE, ID, and DoS) are used to categorize and prioritize the reported CVEs.

To demonstrate the analytical capabilities of our framework, Table  \ref{tab:benchmark-filtering} details the percentage of CVEs filtered for each OS, which ranges from $0.6\%$ for AAOS 34 up to $59.8\%$ for AutoSD, alongside a breakdown of CVE classes per OS.

In Table \ref{tab:benchmark-vuln-type} , we leveraged the CVSS Vector Strings (specifically the metrics used to calculate their base scores) to classify each CVE into distinct impact categories: Remote Code Execution (RCE), Local Privilege Escalation (LPE), Information Disclosure (ID, data leakage), Denial of Service (DoS, program disruption), or Unknown when the vector is ambiguous. The only exception is Vanir's results for Android-based OSes, where the vulnerability type is explicitly indicated in their report. It should be noted that Sandbox and Virtual Machine Escapes (SBX/VME) are difficult to isolate relying solely on CVSS vectors, because their scoring metrics closely mirror those of LPEs, they are often reported under the LPE category.

Table \ref{tab:cves-comparison}  presents a comprehensive comparison of total CVEs/EUVDs for each OS, categorized by type (OS or middleware), purpose (safety or non-safety), and their respective certifications, such as ISO/IEC 15408-5 (EAL) and ISO 26262 (ASIL). Finally, Table \ref{tab:os-details}  provides detailed specifications for the tested operating systems.

To the best of our knowledge, VERA is the first framework to propose automated report parsing, filtering, and sorting (including specific CVE impact classes such as RCE, LPE, ID, and DoS) across multiple vulnerability scanners, coupled with a dynamic exploit finder. This approach provides a practical, unified evaluation methodology specifically tailored for the operating systems and middleware used in next-generation vehicles.

A CVE shall be seen as a potential attack vector, rather than a vulnerability that can be exploited with certainty.
The more CVEs/EUVDs one OS has, the more it has potential attack vectors.
We also tested Ubuntu 22.04 LTS and 20.04 LTS, as well as Android, because numerous OS rely on them. 

Libraries related to archive handling and data encoding (e.g., \texttt{libarchive}, \texttt{zlib}), image processing (e.g., \texttt{libpng}, \texttt{libjpeg-turbo}), XML parsing (e.g., \texttt{libxml2}), database management (e.g., \texttt{libsqlite}), and SSL/TLS (e.g., \texttt{libssl}) are extensively used by both the system and applications and therefore constitute attractive targets for exploitation. In addition, some applications rely on custom embedded libraries, which may also introduce exploitable vulnerabilities.

Initial automated scans revealed multiple critical vulnerabilities (CVEs and EUVDs). To demonstrate feasibility, we successfully exploited two of these vulnerabilities in a controlled environment. (1) We successfully exploited \texttt{CVE-2022-35737} affecting SQLite library running under Android Automotive OS,\footnote{See the following videocapture for validation purposes: \url{https://github.com/EternalDreamer01/vera/blob/main/demo/cve-sqlite.mp4}} but it could be exploited in any application using SQLite database engine subject to the 2GB memory constraint.
Further works could search for vulnerabilities affecting other libraries, and their exploitability in known applications.
(2) we  implemented the de-association attack (Zelle et al. \cite{daniel2021someip})\footnote{A de-association attack targets the service discovery mechanism in SOME/IP by convincing clients that a legitimate service is no longer available. In this attack, an attacker eavesdrops on broadcast service discovery messages (FindService) to identify clients interested in a specific service. The attacker then sends spoofed StopOffer messages to these clients, claiming the service has been discontinued.
} against AAOS, AutoSD, and TeslaOS, targeting a SOME/IP service hosted in a container on the same network. 
On AAOS, the attack was unsuccessful, potentially due to security mechanisms that randomize port numbers. 
On AutoSD and TeslaOS the attack succeeded, allowing us to disconnect the target container from the legitimate service.\footnote{The following videocapture shows the feasibility and success of the attack: \url{https://github.com/EternalDreamer01/vera/blob/main/demo/attack-deassociation.mp4}}

The testbed architecture used for these experiments deployed the target subsystem inside one or two Docker containers (one container for a standalone service or two containers when a client/server pair was required for SOME/IP), while the attacker ran in a separate Docker container attached to the same Docker network (bridge). All network traffic between attacker and target was confined to the Docker network to reproduce a realistic intra-host lateral movement scenario.

Subsequently, to assess realistic attack progression, we executed two multistage scenarios that chained initial exploitation with lateral movement; both scenarios succeeded.

\paragraph{Scenario A -- Remote compromise via the infotainment application store (safety threat)} a victim installs an application obtained through the vehicle’s infotainment application store that embeds a covert backdoor.  Upon execution, the backdoor establishes persistent access to the infotainment unit, thereby enabling attackers to remotely compromise a large number of vehicles. This scenario highlights the potential for catastrophic, large-scale attacks~\cite{he2018security}.

\paragraph{Scenario B -- Remote compromise via infotainment for personal data collection}
an attacker exploits a known vulnerability in the infotainment system to achieve remote code execution (RCE).
From this initial foothold, the attacker might want leverage a local privilege escalation (LPE) vulnerability to grant themself more privileges and capabilities.
This step is not always necessary, depending on the initial exploited application and objectives.
The attacker can then collect any available data such as addresses, contacts, and messages.
A likely consequence of this attack is that an adversary who attains control of the infotainment unit can pivot via a vulnerable gateway ECU to the vehicle's internal CAN/Automotive Ethernet (AE) network, enabling the injection of forged  frames and the issuance of malicious commands to safety-critical ECUs (e.g., braking or steering controllers) \cite{huq2024cybersecurity}\cite{uddin2023systematic}.

Our threat model evaluates specific attack scenarios by adopting the Threat Analysis and Risk Assessment (TARA) methodology from the ISO/SAE 21434 standard \cite{iso21434}. 
Table \ref{tab:attack_scenario} describes the discussed scenarios mapping their tactics, techniques, and potential impacts. 
The specific tactics and techniques are derived from the Automotive Threat Matrix (ATM)\footnote{The Auto-ISAC ATM \url{https://atm.automotiveisac.com}, is a valuable resource for analyzing complex attacks in CAVs. Inspired by the MITRE ATT\&CK framework, the ATM defines 11 tactics representing attacker objectives and 85 techniques detailing specific attack methods. These tactics include Initial Access, Execution, Persistence, Privilege Escalation, Defense Evasion, Credential Access, Discovery, Lateral Movement, Collection, Command and Control, Exfiltration, Manipulate Environment, and Affect Vehicle Function. The ATM is particularly useful for visualizing multi-stage attacks by mapping the sequence of techniques an attacker might employ}~\cite{AutomotiveThreatMatrix}.
Furthermore, we evaluate the consequences of these scenarios using the four impact categories defined by TARA: safety, financial, operational, and privacy, with each category rated on a scale from Negligible to Severe.
For Scenario A, we estimate:
\begin{itemize}
	\item Severe safety impact: causing accidents via braking or acceleration manipulation;
	\item Major financial impact: costs associated with recalls, legal liabilities;
	\item Moderate operational impact: disruption of vehicle functionality, recalls or repairs;
	\item Negligible privacy impact: limited data exposure.
\end{itemize}
For Scenario B, we estimate:
\begin{itemize}
	\item Moderate safety impact: location tracking for physical harm (indirect safety threat);
	\item Moderate financial impact: fines related to data leakage;
	\item Moderate operational impact: reduced functionality of infotainment systems, recalls or repairs;
	\item Severe privacy impact: personal data exposure.
\end{itemize}

Table \ref{tab:feasibility_factors} and \ref{tab:feasibility_scale} describe the values and scales used for our calculations in Table \ref{tab:attack_feasibility}; the sum of the five factors (Table \ref{tab:feasibility_factors}) describe the attack potential/feasibility (cf. Table \ref{tab:feasibility_scale}). We considered the overall feasibility to be determined by the lowest feasibility among the techniques composing the scenario.

\begin{figure*}[t!]
	\centering
	\subfloat[]{\includegraphics[width=.99\textwidth]{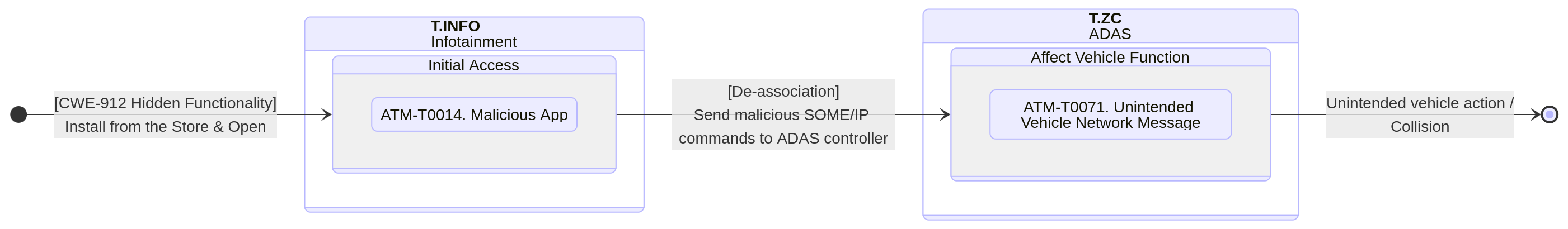}}
	\hspace{0.2cm}
	\subfloat[]{\includegraphics[width=.99\textwidth]{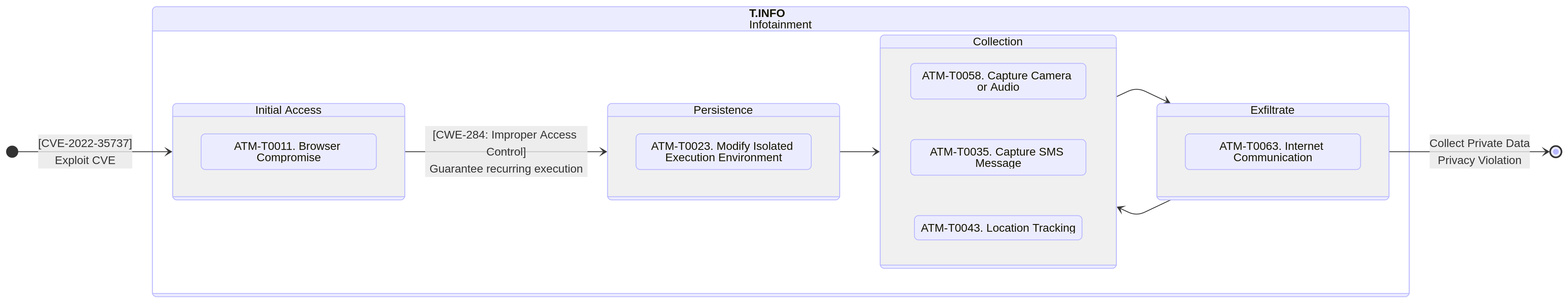}}
	\caption{Killchain state machines for the paths of scenarios A and B.}
	\label{fig:killchains}
\end{figure*}

Table \ref{tab:risk_matrix} is the risk matrix: to determine the risk with a given impact (evaluated on Table \ref{tab:attack_scenario}) and feasibility (overall feasibility calculated on Table \ref{tab:attack_feasibility}). This risk matrix is used to assess the risk of our attack scenarios on Table \ref{tab:attack_scenario_risks}. We consider the overall risk to be determined by the highest risk among the different risk of impact categories.
Figure \ref{fig:killchains} illustrates the killchain state machines for the two attack scenarios.

\section{Discussion}
\label{sec:discussion}

\paragraph{Key findings}

The comprehensive scanning of embedded operating systems and middleware reveals a substantial and heterogeneous vulnerability landscape across the examined platforms. The results presented in Section~\ref{sec:experimental} validate our claims, i.e., that vulnerability prevalence exhibits significant variance depending on the OS/middleware under examination, with severity distributions ranging from primarily low-risk findings to critically concerning threat vectors. 

Among the non-Android/Yocto platforms, Eclipse S-CORE exhibits the lowest vulnerability count (8 total), with no critical findings and a moderate distribution across low and medium severity levels. This profile aligns with its specialized design purpose and potentially more controlled dependency footprint. Conversely, platforms such as ROS2 (281 total CVEs/EUVDs), TeslaOS (243), and AutoSD/RHIVOS (129)  show substantially higher vulnerability burdens. ROS2 presents an exceptionally high-risk profile with 28 critical-severity vulnerabilities, significantly surpassing other platforms in this severity category. This concentration of critical vulnerabilities in ROS2 represents an attack surface that warrants immediate attention, particularly given ROS2's adoption in CAVs but also in robotics and autonomous systems research.
Regarding the Android and Yocto-based platforms, the number of reported CVEs is significantly higher than for non-Android/Yocto systems. Among these, AAOS 34 exhibits the lowest vulnerability count (159 CVEs), whereas AGL presents the highest (1203 CVEs).

The CVSS severity stratification reveals distinct vulnerability patterns across platform categories. Traditional operating systems demonstrate moderate vulnerability loads: Ubuntu 22.04 and 20.04 (both with 22 total CVEs/EUVDs) show well-distributed low and medium vulnerabilities with minimal critical exposures. However, safety-critical platforms certified under ISO/IEC 15408-5 (EAL) or assessed under ISO 26262 (ASIL) are not immune. For instance, VxWorks 7 (EAL 6+) and QNX Neutrino (EAL 4+), exhibit significant totals of 33 and 56 CVEs/EUVDs respectively, that include numerous of high-severity CVSS, demonstrating that certification lowers but does not eliminate observable attack surface.
Finally, Android and Yocto-based platforms exhibit a concerningly high number of critical and high-severity CVSS vulnerabilities  (e.g., Android 30 accounts for 71 critical and 560 high-rated CVEs). This finding is particularly worrying given that most in-vehicle infotainment systems rely on these operating systems (cf. Table \ref{tab:mapping_cars}). Moreover, the elevated prevalence of KEVs and publicly available exploits further underscores the inherent fragility and large attack surface of these platforms, as well as the maturity of the adversarial ecosystem targeting them.

Finally, regarding the comparison of our tool VERA with existing scanners, Table \ref{tab:cves-comparison} also reports results produced by other tools. For non-Android and non-Yocto operating systems we compared VERA against Grype\textcolor{black}{, Vanir, \texttt{cve-check} and CBT}. Trivy’s outputs proved highly inconsistent. More precisely, \textcolor{black}{for CentOS-based (e.g., AutoSD) and Ubuntu-based images (e.g., QNX and TeslaOS) it only reported Go-related CVEs, it returned no findings on Alpine-based images (e.g., Eclipse S-CORE), and it produced a very large number of false positives ($>$ 1000) for ROS-embedded images}. For these reasons, Table \ref{tab:cves-comparison} presents only the Grype results.

We emphasize that VERA’s objective differs from that of many generic scanners. In other words, rather than exhaustively listing every CVE, VERA applies systematic filtering and targeted analysis to surface vulnerabilities that are plausibly exploitable in next-generation vehicle contexts. 

Therefore, after this filtering and follow-on investigation, VERA typically reports fewer CVEs than Grype, deliberately trading breadth for higher signal-to-noise so that analysts and decision-makers can concentrate on true, vehicle-relevant priorities. 
By prioritizing actionable, context-specific vulnerabilities, VERA reduces remediation noise and enables more effective, risk-based allocation of security resources.

\paragraph{Interpretation}

First, raw CVE/EUVD counts should be interpreted as potential attack surface rather than proof of immediate exploitability. Notice that a CVE denotes a known weakness that could be used as an attack vector under the right conditions. However, the exploitation depends on configuration, deployed components, and presence of vulnerable code paths. Hence, the higher counts seen for some OSes/middleware indicate a larger pool of attack vectors that defenders must consider, not that those systems are necessarily trivially exploitable in their stock configurations. This nuance is central to risk assessment and triage. 

Second, the distribution of CVEs correlates with two practical realities. One is \textit{codebase and dependency surface}, maning that OSes and middleware that carry many userland packages, or third-party libraries (archive/image/XML/DB/SSL libraries called out in our evaluation), naturally surface more CVEs. 
The other is \textit{visibility}, where open platforms with richer ecosystems and more public scrutiny (e.g., Android/AGL) tend to have more reported CVEs simply because more researchers, vendors, and scanners examine them. Both effects are visible in our results. 

Third, certification (EAL/ASIL) and perceived safety posture do not eliminate the problem. Indeed, certified kernels and RTOSes still show non-zero vulnerability counts (Table \ref{tab:cves-comparison}). 
Certification can constrain attack surface and enforce development controls, but it is not a substitute for continuous discovery/patching of flaws in the large software stack that surrounds an RTOS in an automotive product.

\paragraph{Limitations}

We note some limitations in our work. First, vulnerability assessment was performed on Dockerized environments to ensure reproducibility and isolation.
While this setup enables systematic comparison, it does not fully capture the complexity of in-vehicle deployments, including vendor-specific kernel builds and configurations, proprietary implementations, firmware interactions, and hardware-dependent behavior. As a result, some vulnerabilities that depend on platform-specific integrations or runtime conditions may be misrepresented \cite{Evaluating_Container_Security_and_Reproducibility_in_Research_Software_Engineering}.
Second, we rely on CVSS scores to characterize vulnerability severity. Although widely adopted, CVSS reflects intrinsic vulnerability properties and does not account for deployment context, available mitigations, or exploit feasibility, which may lead to over or under-estimation of real-world risk.
Third, the presence of a potential vulnerability does not imply practical exploitability. Many reported issues require specific compilation options, configurations, or operational conditions that may not be satisfied in actual automotive systems. A precise assessment of exploitability would require access to real-world configurations and proprietary settings, which are outside the scope of this study.

This gap described by the second and third limits is well illustrated by our de-association case study. The same reported weakness was exploitable on AutoSD and TeslaOS but failed against AAOS, showing that identical severity ratings may lead to divergent operational outcomes depending on platform configuration and runtime defenses. To reduce such mismatches,
VERA can help through semi-automated exploitability checks (especially for symbols/functions presence and calls).
By contrast, the other comparative scanners report package-level matches without this additional filtering.


\section{Related work}
\label{sec:rw}
\begin{table*}[ht]
	\newcolumntype{$}{>{\global\let\currentrowstyle\relax}}
	\newcolumntype{^}{>{\currentrowstyle}}
	\newcommand{\rowstyle}[1]{\gdef\currentrowstyle{#1}%
	#1\ignorespaces
	}
\centering
\scriptsize
\caption{Comparison of related works in automotive cybersecurity and vulnerability scanning}
\label{tab:related_works}
\begin{tabular}{l l c c c c c p{5cm}}
\toprule
\textbf{Ref.} & \textbf{Primary Focus} & \textbf{SDV} & \textbf{CAV} & \textbf{Vuln. Scan} & \textbf{Offline} & \textbf{Scalable} & \textbf{Key Contribution / Limitation} \\
\midrule
Kifor, Popescu \cite{kifor2024automotive} & Cybersecurity frameworks & \filledcircle & \filledcircle & \semicircle & \emptycircle & \semicircle & \textit{+} Surveys SDV security frameworks. \newline \textit{--} Limited focus, no offline support. \\
Huq \cite{huq2024cybersecurity} & SDV vulnerabilities & \filledcircle & \filledcircle & \emptycircle & \emptycircle & \emptycircle & \textit{+} Identifies SDV risks (infotainment, OTA). \newline \textit{--} No focus on vulnerability scanning or scalability. \\
Uddin et al. \cite{uddin2023systematic} & Cybersecurity trends & \semicircle & \filledcircle & \emptycircle & \emptycircle & \emptycircle & \textit{+} Reviews CAV attack vectors. \newline \textit{--} No specific SDV or scanning focus, not scalable. \\
Eiza, Ni \cite{eiza2017cybersecurity} & Connected vehicle threats & \semicircle & \filledcircle & \emptycircle & \emptycircle & \emptycircle & \textit{+} Identifies connectivity risks. \newline \textit{--} Limited SDV focus, no scanning or scalability. \\
Wang, Zhang \cite{wang2021systematic} & TARA for CAVs & \semicircle & \filledcircle & \emptycircle & \emptycircle & \semicircle & \textit{+} TARA-based risk assessment. \newline \textit{--} No vulnerability scanning, limited SDV focus. \\
Jayarathne et al.\cite{jayarathne2024simulation} & Simulation-based risk & \semicircle & \filledcircle & \emptycircle & \emptycircle & \semicircle & \textit{+} Simulation for CAV risks. \newline \textit{--} No scanning, limited SDV focus. \\
Jeong et al.\cite{jeong2022infotainment} & AGL infotainment security & \filledcircle & \filledcircle & \emptycircle & \emptycircle & \emptycircle & \textit{+} Demonstrates AGL attacks. \newline \textit{--} Platform-specific, no scanning or scalability. \\
Gong et al.\cite{gong2023protection} & Android Automotive security & \filledcircle & \filledcircle & \emptycircle & \emptycircle & \emptycircle & \textit{+} Lightweight protection for Android. \newline \textit{--} Platform-specific, no scanning or scalability. \\
Buczak, Guven \cite{buczak2022survey} & Container vuln. scanning & \emptycircle & \emptycircle & \filledcircle & \semicircle & \semicircle & \textit{+} Evaluates scanner limitations. \newline \textit{--} Not automotive-specific, partial offline support. \\
Doan, Jung \cite{doan2022dav} & Dockerfile vuln. scanning & \emptycircle & \emptycircle & \filledcircle & \semicircle & \semicircle & \textit{+} Improves container scanning. \newline \textit{--} Not automotive-specific, partial offline support. \\
Zhang et al.\cite{zhang2023libraries} & Open Source Libraries and CVEs & \filledcircle & \filledcircle & \semicircle & \emptycircle & \semicircle & \textit{+} Identifies libraries and associated CVEs \newline \textit{--} Limited review of possible CVEs. \\
De Vincenzi et al. \cite{Contextualizing_security_and_privacy_of_software_defined_vehicles_State_of_the_art_and_industry_perspectives} & Security and privacy & \filledcircle & \filledcircle & \emptycircle & \emptycircle & \emptycircle & \textit{+} Review security and privacy concerns \newline \textit{--} No vulnerability scanning. \\
Jeong et al.\cite{jeong2023infotainment} & Impact and implications of IVI & \filledcircle & \filledcircle & \emptycircle & \emptycircle & \emptycircle & \textit{+} Security and privacy concerns implied by IVI \newline \textit{--} No vulnerability scanning. \\
This work & Security evaluation for SDV & \filledcircle & \filledcircle & \filledcircle & \semicircle & \filledcircle & \textit{+} Practical security evaluation on SDV OSes \newline \textit{--} SBX/VME not fully addressed, no real-world testing \\

\bottomrule
\end{tabular}

\vspace{0.25cm}
\parbox{\linewidth}{\scriptsize
\textbf{Legend:} 
\filledcircle\ = Explicit focus, 
\semicircle\ = Partial or likely focus, 
\emptycircle\ = Not addressed. \\
\textbf{SDV:} Focus on software-defined vehicles or POSIX-based systems. 
\textbf{CAV:} Connected and Autonomous Vehicles. 
\textbf{Vuln. Scan:} Vulnerability scanning tools or methods. 
\textbf{Offline:} Support for offline analysis. 
\textbf{Scalable:} Support for scalable, multi-product analysis.
}
\end{table*}

Our work builds upon research in automotive security, SDV architectures, and vulnerability scanning. This section surveys the most relevant contributions in these areas and highlights the gaps our work aims to address.

Table \ref{tab:related_works} summarizes the related works with respect to key criteria relevant to VERA's focus on automotive cybersecurity and vulnerability scanning.

\paragraph{Automotive security analysis}

Recent studies have advanced the understanding of automotive cybersecurity, particularly for Connected and Automated Vehicles (CAVs). Eiza et al. \cite{eiza2017cybersecurity} reviewed cybersecurity threats in connected vehicles, identifying vulnerabilities in connectivity interfaces such as cellular and V2X systems. Uddin et al. \cite{uddin2023systematic} provided a systematic review of recent trends, highlighting remote attack vectors and the increasing complexity of vehicle software stacks. Wang et al. \cite{wang2021systematic} proposed a TARA-based risk assessment aligned with ISO/SAE 21434, focusing on systematic threat modeling for CAVs. Similarly, Jayarathne et al. \cite{jayarathne2024simulation} introduced simulation-based risk assessment for CAVs, emphasizing practical evaluation of attack scenarios. Pitchamaini et al. \cite{Systematic_Risk_Analysis_of_Multi_Stage_Attacks_in_Zonal_Automotive_E_E_Architecture}  proposed an approach that integrate the Automotive Threat Matrix (ATM) with the ISO/SAE 21434 TARA process to systematically construct and analyze attack paths. In the same context, Benyahya et. al  \cite{TARA_2_0_for_Connected_and_Automated_Vehicles} proposed TARA 2.0, a TARA based framework that focuses on privacy.

While prior studies and existing vulnerability scanners establish a strong foundation, they cannot be directly extended to the POSIX-based OSes used in modern vehicles. The primary limitation of these generic tools is their lack of automotive context: they are too broad and produce an unmanageable amount of noise (mainly false positives) when applied to the highly customized, stripped-down software stacks typical of automotive environments. As demonstrated in our results (Tables \ref{tab:cve-cvss-benchmark} and \ref{tab:cve-cvss-benchmark-2}), generic scanners fail to account for automotive-specific package filtering, leading to inflated vulnerability counts.

\paragraph{Security of SDVs}

The shift to SDVs has introduced new security challenges due to increased software complexity and connectivity. Kifor et al. \cite{kifor2024automotive} provided a comprehensive survey of automotive cybersecurity frameworks, including those applicable to SDVs, emphasizing the need for robust testing and monitoring. Huq et al. \cite{huq2024cybersecurity} discussed emerging risks in SDVs, such as vulnerabilities in infotainment systems and OTA updates, which align with our threat model. \textit{Sghaier et al.} \cite{Advancing_Security_in_Software_Defined_Vehicles_A_Comprehensive_Survey_and_Taxonomy} provided a comprehensive systemization of knowledge  on SDVs, mapping the ecosystem and enabling technologies, identifying principal cyberattack entry points that arise from their architectural and operational characteristics, and proposing a taxonomy of SDV-specific attacks.

Specific platforms used in SDVs have also been studied. Jeong et al. \cite{jeong2022infotainment} analyzed the security of Automotive Grade Linux (AGL), demonstrating practical attacks on infotainment systems. Gong et al. \cite{gong2023protection} proposed a lightweight protection mechanism for vehicle control functions on Android Automotive OS. These studies highlight the critical role of the OS layer in SDV security but focus on single platforms. In contrast, VERA is designed for scalable analysis across multiple platforms and products, addressing a broader range of SDV environments.

\textit{Zhang et al.} \cite{zhang2023libraries} analysed open source components from automotive firmware and compared them with commonly used components in general-purpose operating systems. Their study reports an average of 79.8 open source components per firmware image, representing roughly 16\% of binary files. 
They further note that some open source artifacts are associated with CVEs. However, the analysis does not employ automated vulnerability-scanning tools. Instead, the authors map library versions to CVE entries listed at NVD
 and enumerate the matches. Consequently, the study is primarily descriptive rather than providing a security-driven vulnerability assessment. More precisely, the study's primary focus is quantifying and characterizing the presence of open source components in automotive platforms rather than systematically assessing their security posture.

\paragraph{Privacy concerns in SDVs}
Data collection from vehicles is long established. \textit{De Vincenzi et al.}~\cite{Contextualizing_security_and_privacy_of_software_defined_vehicles_State_of_the_art_and_industry_perspectives} provided a comprehensive analysis of the security and privacy challenges associated with SDVs, emphasizing a range of architectural and operational  security and privacy risks. 
Mainly, they observe that privacy is frequently treated as secondary to security in SDV research and practice. This oversight leaves privacy as an under-recognized and exploitable attack surface.

Jeong et al. \cite{jeong2023infotainment} highlights that an attacker can track location, obtain phonebook entries, recent call logs, and text messages from Bluetooth-paired smartphones, make a phone call or send a message to any recipient.
In 2023, the Mozilla foundation exposed serious privacy violations after researching the privacy policies of multiple vehicles manufacturers \cite{Mozilla_Privacy_Review}.
Automobiles from manufacturers such as Nissan, Volkswagen, Toyota and others have been shown to gather private information about their drivers, including sexual activity, immigration status, ethnicity, facial expressions, weight, health and genetic information, and where you drive.
These highly sensitive personal data are routinely shared with third parties, including advertisers and data brokers, yet manufacturers frequently provide limited transparency regarding what is collected and how it is disseminated \cite{Mozilla_Privacy_Review}. As vehicles become persistently connected, these data streams also expand the attack surface, increasing exposure to unauthorized access, profiling, and other forms of misuse and privacy violation.

\paragraph{Vulnerability scanning tools}

Several open source options, such as Trivy, Grype, and OSV-Scanner, are widely used to detect CVEs in software dependencies and container images \cite{buczak2022survey}. They have limitations in the automotive context \textit{Buczak et al.} \cite{buczak2022survey}, including inconsistencies in container vulnerability scanning and reliance on online databases. \textit{Doan et al.} \cite{doan2022dav} proposed DAVS, a method to improve container scanning by analyzing Dockerfiles, but it does not address the need for offline-first analysis of pre-compiled artifacts.

In this work we propose VERA,
(1) a suite of tools, that filters, sorts and prioritizes CVEs. (2) it provides functionalities to scan layered/Open Container Initiative (OCI) container images or Android emulator directly, and (3) provides guidance to assess exploitability, retrieve information on a specific CVE and discover online exploits.
Unlike penetration testing frameworks like those discussed in \cite{kifor2024automotive}, which focus on black-box testing, VERA provides a systematic semi-white-box approach to identify known vulnerabilities through packages scanners. It also handles the scanning process through CBT
(as in black-box settings), and confirm the presence of some CVEs through integrated reverse-engineering tools. Hence, it enables a proactive, in-depth pentesting strategy.

\section{Conclusion}
\label{sec:conclusion}

Modern vehicles, denoted in our work as Software-defined vehicles (SDVs), are increasingly adopting Portable Operating System Interface (POSIX)-compatible platforms to support new features.  Motivated by a simple but consequential observation (POSIX-compatible operating systems and products have long accumulated large numbers of CVEs) we have provided a comprehensive security analysis for SDVs, focusing on software vulnerabilities. We have also presented VERA, a custom vulnerability assessment solution tailored to efficiently discover vulnerabilities on different operating systems, within a dockerized development environment to evaluate exploitability issues. VERA is conceived as a suite for reproducible vulnerability analysis and prioritization, which  standardizes and compresses scanner outputs, resolves unknown CVSS and EPSS entries. 
We also contributed to the CVE Binary Tool (CBT) project, to better support Android's binaries, and include the results in this paper.

As perspectives for future work, we envision refining the vulnerability assessment by incorporating real-world system configurations and usage patterns to better assess the presence and the practical exploitability of identified vulnerabilities. Scanning more automotive operating systems like PikeOS, INTEGRITY, NVIDIA DRIVE OS, etc. would also be interesting. We might focus on a specific OS, like an Android infotainment system as it could be a potential entry point for an attacker, to go deeper in the analysis of vulnerabilities and their exploitability.
We could also attempt to implement fully automated vulnerability exploitability testing, extracting the conditions from the CVE data especially the vulnerable symbols.

\section*{Acknowledgment}

The work presented in this paper was conducted within the framework of the Horizon Europe AI4CCAM project (grant agreement 101076911), and the Carnot project CARNOT 2025 -TSN-D6/AMI Citroen - véhicule autonome of Télécom SudParis.

\bibliographystyle{unsrt}
\bibliography{references}

\end{document}